%% file: main.tex
\documentclass[screen, camera, acmsmall, natbib=false]{acmart}

\usepackage{xspace}
\usepackage{mathpartir}
\usepackage{mathtools}
\usepackage{amsmath,amsfonts}
\usepackage{listings}
\usepackage{subcaption}
\usepackage{caption}
\usepackage{wrapfig}
\usepackage{xcolor}
\newcommand{\mathcolorbox}[2]{\setlength{\fboxsep}{0.25pt}\colorbox{#1}{$\displaystyle #2$}}

\newcounter{redcnt}
\AtBeginDocument{%
  }

\setcopyright{none}
\renewcommand\footnotetextcopyrightpermission[1]{}

\RequirePackage[
  datamodel=acmdatamodel,
  style=numeric,
  ]{biblatex}

\bibliography{bib,dbp,amal}

\newcommand{\rwasm}{RichWasm\xspace}
\newcommand{\lthree}{$L^3$\xspace}
\newcommand{\ml}{ML\xspace}

\input{defs}
\begin{document}

\title{\rwasm: Bringing Safe, Fine-Grained, Shared-Memory Interoperability Down
  to WebAssembly}

\author{Zoe Paraskevopoulou}

\affiliation{
  \institution{Northeastern University}
  \country{USA}
}

\author{Michael Fitzgibbons}

\affiliation{
  \institution{Northeastern University}
  \country{USA}
}

\author{Noble Mushtak}

\affiliation{
  \institution{Northeastern University}
  \country{USA}
}

\author{Michelle Thalakottur}

\affiliation{
  \institution{Northeastern University}
  \country{USA}
}

\author{Jose Sulaiman Manzur}

\affiliation{
  \institution{Northeastern University}
  \country{USA}
}

\author{Amal Ahmed}

\affiliation{
  \institution{Northeastern University}
  \country{USA}
}

\begin{abstract}
  Safe, shared-memory interoperability between languages with
  different type systems and memory-safety guarantees is an intricate problem as 
  crossing language boundaries may result in memory-safety violations.
  In this paper, we present RichWasm, a novel richly typed
  intermediate language designed to serve as a compilation target for
  typed high-level languages with different memory-safety guarantees.
  RichWasm is based on WebAssembly and enables safe shared-memory
  interoperability by incorporating a variety of type features
  that support fine-grained memory ownership and sharing.
  RichWasm is rich enough to serve as a typed compilation target for
  both typed garbage-collected languages and languages with an
  ownership-based type system and manually managed memory.
  We demonstrate this by providing compilers from core ML and
  \lthree, a type-safe language with strong updates~\cite{l3}, to RichWasm.
  RichWasm is compiled to regular Wasm, \mbox{allowing for use in existing environments.  We formalize RichWasm in Coq and prove type safety.}
\end{abstract}

%% \begin{CCSXML}
%% <ccs2012>
%%  <concept>
%%   <concept_id>10010520.10010553.10010562</concept_id>
%%   <concept_desc>Computer systems organization~Embedded systems</concept_desc

\keywords{WebAssembly, memory ownership, memory sharing, capability types}

\maketitle

\input{intro}
\input{overview}

\input{dynamics}
\input{statics}
\input{compilers}

\input{lowering}
\input{related}
\input{discussion}

\printbibliography

%\section{Appendix Section}

\end{document}

%% file: defs.tex
\newcommand {\Wkey}[1]{\boldsymbol{\mathsf{#1}}}
\newcommand {\Wmath}[1]{#1} %backwards comp
\newcommand {\Wsf}[1]{{\mathsf{#1}}}
\newcommand {\Wbf}[1]{{\mathbf{#1}}}

\newcommand {\bnflabel}[1]{\mbox{\textit{#1}}}
\newcommand {\bnfalt}{\mathrel{\bf \mid }}
\newcommand {\bnfaltend}{\mathrel{\bf \,\mid}}
\newcommand {\bnfdef}{\mathrel{\bf::=}}

\newcommand {\deftable}[1]{\[\begin{array}{l l l l} #1 \end{array}\]}
\newcommand {\defline}[3]{\bnflabel{#1} & #2 &\!\!\!\! \bnfdef \!\!\!\! & \!\! #3 \\}
\newcommand {\defnewline}[1]{ & & & \!\! #1 \\}

\newcommand{\many}[1]{{#1}^{*}}
\newcommand{\manyn}[2]{#1^{#2}}
\newcommand{\mhl}[1]{\mathcolorbox{black!15}{#1}}

\newcommand {\Wloc}{\Wmath{\ell}}
\newcommand {\Wloclin}{\Wmath{\ell}_{\Wlin}}
\newcommand {\Wlocunr}{\Wmath{\ell}_{\Wunr}}
\newcommand {\Wlocvar}{\rho}

\newcommand {\Wsz}{\mathit{sz}}
\newcommand {\Whv}{\mathit{hv}}
\newcommand {\Wszvar}{\sigma}

\newcommand {\Wqual}{q}
\newcommand {\Wqualvar}{\delta}

\newcommand {\Wsign}{sx}

\newcommand {\Wrw}{\pi}
\newcommand {\Wrwrw}{\Wkey{rw}}
\newcommand {\Wrwr}{\Wkey{r}}

\newcommand {\Wnp}{\mathit{np}}
\newcommand {\Wpty}{p}
\newcommand {\Wptyvar}{\alpha}

\newcommand {\Wtau}{\tau}
\newcommand {\Wty}{\tau}

\newcommand {\Whty}{\psi}

\newcommand {\Waty}{\mathit{tf}}
\newcommand {\Wglob}{\mathit{tg}}

\newcommand {\Wk}{\kappa}
\newcommand {\Wfty}{\chi}

\newcommand {\Wtyless}{\lesssim}
\newcommand {\Wszless}{\leq}
\newcommand {\Wqualless}{\preceq}

\newcommand {\Wleff}{\many{(i, \Wty)}}

\newcommand {\Wunr}{\Wkey{unr}}
\newcommand {\Wlin}{\Wkey{lin}}

\newcommand {\WPunit}{\Wkey{unit}}
\newcommand {\WPgroup}[1]{(#1)}
\newcommand {\WPcoderef}[1]{\Wkey{coderef}~#1}
\newcommand {\WPrec}[3]{\Wkey{rec}\ #1 \Wqualless #2.\ #3}
\newcommand {\WPptr}[1]{\Wkey{ptr}\ #1}
\newcommand {\WPexists}[2]{\exists #1.\ #2}
\newcommand {\WPcap}[3]{\Wkey{cap}\ #1\ #2\ #3}
\newcommand {\WPref}[3]{\Wkey{ref}\ #1\ #2\ #3}
\newcommand {\WPown}[1]{\Wkey{own}\ #1}

\newcommand {\WHvariant}[1]{(\Wkey{variant}\ #1)}
\newcommand {\WHstruct}[1]{(\Wkey{struct}\ #1)}
\newcommand {\WHarray}[1]{(\Wkey{array}\ #1)}
\newcommand {\WHexists}[4]{(\exists\ #1 \Wqualless #2 \Wtyless #3.\ #4)}

\newcommand{\WVunit}{\Wkey{()}}
\newcommand{\WVcoderef}[3]{\Wkey{coderef}~#1~#2~#3}
\newcommand{\WVfold}[1]{\Wkey{fold}~#1}
\newcommand{\WVref}[1]{\Wkey{ref}~#1}
\newcommand{\WVptr}[1]{\Wkey{ptr}~#1}
\newcommand{\WVcap}{\Wkey{cap}}
\newcommand{\WVown}{\Wkey{own}}
\newcommand{\WVmempack}[2]{\Wkey{mempack}~#1~#2}

\newcommand {\WHVvariant}[2]{(\Wkey{variant}~#1~#2)}
\newcommand {\WHVstruct}[1]{(\Wkey{struct}~#1)}
\newcommand {\WHVarray}[2]{(\Wkey{array}~#1~#2)}
\newcommand {\WHVpack}[3]{(\Wkey{pack}~ #1 ~#2 ~#3)}

\newcommand {\STEITeqz}{\Wkey{eqz}}
\newcommand {\STEIReq}{\Wkey{eq}}
\newcommand {\STEIRne}{\Wkey{ne}}
\newcommand {\STEIRlt}[1]{\Wkey{lt}\ #1}
\newcommand {\STEIRgt}[1]{\Wkey{gt}\ #1}
\newcommand {\STEIRle}[1]{\Wkey{le}\ #1}
\newcommand {\STEIRge}[1]{\Wkey{ge}\ #1}
\newcommand {\STEFReq}{\Wkey{eq}}
\newcommand {\STEFRne}{\Wkey{ne}}
\newcommand {\STEFRlt}{\Wkey{lt}}
\newcommand {\STEFRgt}{\Wkey{gt}}
\newcommand {\STEFRle}{\Wkey{le}}
\newcommand {\STEFRge}{\Wkey{ge}}
\newcommand{\MST}[1]{#1} %% backwards compat
\newcommand{\Wunreach}{\Wkey{unreachable}}
\newcommand{\Wnop}{\Wkey{nop}}
\newcommand{\Wdrop}{\Wkey{drop}}
\newcommand {\Wselect}{\Wkey{select}}
\newcommand {\Wblock}[3]{\MST{\Wkey{block}\ #1\ #2\ #3\ \Wkey{end}}}
\newcommand {\Wloop}[2]{\MST{\Wkey{loop}\ #1\ #2\ \Wkey{end}}}
\newcommand {\Wite}[4]{\MST{\Wkey{if}\ #1\ #2\ #3\ \Wkey{else}\ #4\ \Wkey{end}}}
\newcommand {\Wbr}[1]{\MST{\Wkey{br}\ #1}}
\newcommand {\Wbrif}[1]{\MST{\Wkey{br\_if}\ #1}}
\newcommand {\Wbrtable}[2]{\Wkey{br\_table}\ #1\ #2}
\newcommand {\Wreturn}{\Wkey{return}}
\newcommand {\Wgetlocal}[2]{\MST{\Wkey{get\_local}\ #1\ #2}}
\newcommand {\Wsetlocal}[1]{\MST{\Wkey{set\_local}\ #1}}
\newcommand {\Wteelocal}[1]{\MST{\Wkey{tee\_local}\ #1}}
\newcommand {\Wgetglobal}[1]{\MST{\Wkey{get\_global}\ #1}}
\newcommand {\Wsetglobal}[1]{\MST{\Wkey{set\_global}\ #1}}
\newcommand {\Wcoderef}[1]{\MST{\Wkey{coderef}\ #1}}
\newcommand {\Winst}[1]{\MST{\Wkey{inst}\ #1}}
\newcommand {\Wcallindirect}{\Wkey{call\_indirect}}

\newcommand {\Wcall}[2]{\MST{\Wkey{call}\ #1\ #2}}
\newcommand {\Wrecfold}[1]{\MST{\Wkey{rec.fold}\ #1}}
\newcommand {\Wrecunfold}{\Wkey{rec.unfold}}

\newcommand {\Wseqgroup}[2]{\MST{\Wkey{seq.group}\ #1\ #2}}
\newcommand {\Wsequngroup}{\Wkey{seq.ungroup}}

\newcommand {\Wcapsplit}{\Wkey{cap.split}}
\newcommand {\Wcapjoin}{\Wkey{cap.join}}
\newcommand {\Wrefdemote}{\Wkey{ref.demote}}
\newcommand {\Wrefsplit}{\Wkey{ref.split}}
\newcommand {\Wrefjoin}{\Wkey{ref.join}}
\newcommand {\Wmempack}[1]{\MST{\Wkey{mem.pack}\ #1}}
\newcommand {\Wmemunpack}[4]{\Wkey{mem.unpack}\ #1\ #2\ #3.~ #4}
\newcommand {\Wstructmalloc}[2]{\MST{\Wkey{struct.malloc}\ #1\ #2}}
\newcommand {\Wstructfree}{\Wkey{struct.free}}
\newcommand {\Wstructget}[1]{\MST{\Wkey{struct.get}\ #1}}
\newcommand {\Wstructset}[1]{\MST{\Wkey{struct.set}\ #1}}
\newcommand {\Wstructswap}[1]{\MST{\Wkey{struct.swap}\ #1}}
\newcommand {\Wvariantmalloc}[4]{\MST{\Wkey{variant.malloc}\ #1\ #2\ #3\ #4}}
\newcommand {\Wvariantcase}[5]{\MST{\Wkey{variant.case}\ #1\ #2\ #3\ #4\ #5\ \Wkey{end}}}
\newcommand {\Warraymalloc}[1]{\Wkey{array.malloc}\ #1}
\newcommand {\Warrayget}{\Wkey{array.get}}
\newcommand {\Warrayset}{\Wkey{array.set}}
\newcommand {\Warrayfree}{\Wkey{array.free}}
\newcommand {\Wexistpack}[3]{\MST{\Wkey{exist.pack}\ #1\ #2\ #3}}
\newcommand {\Wexistunpack}[6]{\MST{\Wkey{exist.unpack}\ #1\ #2\ #3\ #4\ #5\ #6\ \Wkey{end}}}
\newcommand {\Wqualify}[1]{\MST{\Wkey{qualify}\ #1}}

\newcommand {\Wtrap}{\Wkey{trap}}
\newcommand {\Wcallcl}[2]{\MST{\Wkey{call}\ #1\ #2}}
\newcommand {\Wlabel}[4]{\MST{\Wkey{label}_{#1}\ #2\ \{#3\}\ #4\ \Wkey{end}}}
\newcommand {\Wlocal}[4]{\MST{\Wkey{local}_{#1}\ \{#2;\ #3\}\ #4\ \Wkey{end}}}

\newcommand {\Wmalloc}[3]{\MST{\Wkey{malloc}\ #1\ #2\ #3}}
\newcommand {\Wfree}{\Wkey{free}}

\newcommand {\Wcl}{cl}

\newcommand{\modctx}{M}
\newcommand{\funcctx}{F}
\newcommand{\locctx}{L}
\newcommand{\storetyp}{S}

\newcommand{\typevalid}[2]{#1 \vdash #2 ~ \mathsf{type} }

\newcommand{\qualvalid}[2]{#1 \vdash #2 ~ \mathsf{qual} }

\newcommand{\valtype}[4]{#1; #2 \vdash #3 : #4 }
\newcommand{\heaptype}[4]{#1; #2 \vdash #3 : #4 }
\newcommand{\instrtype}[7]{#1; #2; #3; #4 \vdash #5 : #6 ~|~ #7  }
\newcommand{\instrtypectx}[3]{\instrtype{\storetyp}{\modctx}{\funcctx}{\locctx}{#1}{#2}{#3}}

\newcommand{\conftype}[5]{\vdash_{#1} #2; #3; #4 : #5}
\newcommand{\conftypefull}[6]{#1; #2 \vdash_{#3} #4; #5: #6}

\newcommand{\memlin}{S_{\mathsf{lin}}}
\newcommand{\memunr}{S_{\mathsf{unr}}}
\newcommand{\linempty}{\memlin = \emptyset}

\newcommand{\gethtunr}[2]{\memunr(#1) = #2}

\newcommand{\linsingl}[2]{\memlin = [#1 \mapsto #2]}

\newcommand{\subst}[2]{[#2/#1]}

\newcommand{\meminst}{S_{\mathsf{inst}}}

\newcommand{\typeenv}{F_{\mathsf{type}}}

\newcommand{\flabel}{F_{\mathsf{label}}}

\newcommand{\flinear}{F_{\mathsf{linear}}}

\newcommand{\flocation}{F_{\mathsf{location}}}

\newcommand{\sizeof}[1]{|| #1 ||_{\typeenv}}

\newcommand{\Warrow}[2]{#1 \rightarrow #2}
\newcommand{\Wemptyarrow}[1]{\epsilon \rightarrow #1}
\newcommand{\Wemptyres}[1]{#1 \rightarrow \epsilon}

\newcommand {\Wquallessctx}{\preceq_{\funcctx_{\mathsf{qual}}}}
\newcommand {\Wszlessctx}{\leq_{\funcctx_{\mathsf{size}}}}

\newcommand {\WL}[1]{{L^{#1}}}

\newcommand{\Wred}[8]{#1;#2;#3;#4 \hookrightarrow_{#8} #5;#6;#7}
\newcommand{\Wredi}[7]{#1;#2;#3;#4 \hookrightarrow_{j} #5;#6;#7}

\newcommand{\redrulei}[2]{#1 & \hookrightarrow_i & #2} 
\newcommand{\redrule}[2]{#1 & \hookrightarrow & #2}

\newcommand{\lapp}[1]{\WL{k}[#1]}

\newcommand {\locs}[1]{\Wbf{locs}(#1)}

\newcommand{\sbst}[2]{[#2/#1]}

\newcommand{\sz}[1]{\Wbf{size}(#1)}

\newcommand{\ftyp}{\Warrow{\many{\Wtau_1}}{\many\Wtau_2}}

\newcommand{\appl}{+\kern-1.3ex+\kern0.8ex}

\newcommand{\gethd}{\mathit{get\_hd}}

\newcommand{\nocaps}{\mathit{no\_caps}_{\typeenv}}

\newcommand{\reftolin}{\tt{ref\_to\_lin}}

\newcommand{\negspace}{\vspace{-1ex}}

%% file: intro.tex
  \negspace
\section{Introduction}
\label{sec:intro}

WebAssembly~\cite{wasm} (Wasm) is a portable binary format that enables
almost-native execution speed for a variety of languages.
With around 40 languages compiling to Wasm, the WebAssembly platform
has huge potential to serve as a platform for language interoperability.
This potential has been recognized for some time, but there are two
impediments. The first is that Wasm takes an all-or-nothing approach to sharing
memory. Each Wasm module has its own memory. If it wants to share a data
structure in its memory with another module, that effectively leaves all of 
its memory exposed to the other module, potentially allowing adversarial code to
access and modify arbitrary parts of that memory via pointer arithmetic.
The multiple memories proposal~\cite{wasmmultimemory} addresses this by allowing
modules to have multiple independent memories and only share some of them while
keeping others private. 
A module can thus protect its private data by placing it in a memory that is not
shared, but this mechanism is too coarse-grained, in both a spatial 
and temporal sense: it does not allow fine-grained sharing of only one data
structure in a memory or sharing for only a while and then allowing access to be revoked.  

The second impediment is that Wasm has only low-level types (32- and 64-bit
integers and floats), so when modules from different languages need to
communicate, there is a question of how to exchange high-level values. The
Interface Types proposal~\cite{wasminterface} aimed to address this by
introducing a set of high-level ``interface types'' (e.g., char, list, record,
variant, and more can be encoded) that can be used to communicate across
modules.  
% accompanied by adaptor modules that adapt/instrument regular modules so they
% can use the high-level types to communicate
However, the Interface Types proposal and its successor, the Component Model
proposal~\cite{wasmcomponent}, support only ``shared-nothing'' interoperability,
requiring that values be copied across language boundaries when necessary. This
ensures memory safety but comes with runtime overhead for copying. 
%\amal{I want to say that IT and CM sit above core Wasm and are not part of the
%  Wasm spec but feels like a bit of a digression.}
% By adopting ``shared-nothing'' interoperability, these proposals ensure memory
% safety, but it have to contend with the cost of copying. 

In this paper, we propose \rwasm, a richly typed intermediate language (IL)
based on Web\-Assembly. Unlike existing proposals in the Wasm ecosystem, \rwasm
supports safe, fine-grained, shared-memory interoperability across Wasm modules,
and it is also motivated by a desire to provide a platform atop which language
implementors can easily develop safe FFIs. 
A particular goal is to enable statically detecting memory safety violations
that happen across source-language boundaries and thus cannot be detected by the
source type system.
For instance, if we mix a type-safe garbage-collected language such as OCaml
with a type-safe non-GC'd language such as Rust and allow them to share memory,
violations may include Rust freeing some garbage-collected memory passed to it
from OCaml, or OCaml copying a mutable reference from Rust that Rust considers
uniquely owned.
We wish to allow source-language designers to add inter-language communication
through a simple foreign-function interface that requires no changes to the
existing source type systems and minimal changes to the syntax of their
languages. 
Then the source-language modules are compiled separately to \rwasm modules. Any
potentially problematic interaction between modules will fail to type
check. Thus, type checking at the \rwasm level guarantees cross-module type and
memory safety. 

\paragraph{Overview of \rwasm.} \rwasm supports both a garbage-collected memory
and a manually managed memory so we can compile languages with these very
different memory management strategies to \rwasm and reason about the
highly intricate problems that arise when sharing memory across such languages.  

At the core of \rwasm's type system are capability
types in the style of \lthree~\cite{l3,ahmed07:L3}, which is a linear language with
locations and safe strong updates. A key idea in \lthree was that when we
allocate a new reference and initialize it with a value 
of type $\tau$, we get back an existential package $\exists
\rho. \mathsf{!Ptr~}\rho \otimes \mathsf{Cap~}\rho~\tau$ that says that there
exists some location $\rho$ and we now have an unrestricted (copyable) pointer
to that location and we have a linear capability that tells us the type of value
currently stored at that location. This linear capability is essentially an
ownership token required to access the reference --- the capability must be
provided to read from, write to, or free the location.
In \rwasm, we similarly have \emph{linear} capabilities that represent (allow
access to) uniquely owned memory and support strong updates. But we also have
\emph{unrestricted} capabilities, which represent (allow access to)
garbage-collected memory and are analogous to ML references in that they only
support type-preserving updates. Unlike \lthree, capability types in \rwasm can
provide either read-only access or read-write access.

Formally, \rwasm has a substructural type system, realized via two qualifiers,
linear and unrestricted, that annotate pretypes. In addition to capability
pretypes, \rwasm supports polymorphism, recursive types, variants, structs,
arrays and existential types. Unlike
\lthree, which assumes a reference cell can hold a value of any size, 
\rwasm has a low-level memory model like Wasm's, where each memory is simply a
sequence of bytes and we must allocate data structures in the ``flat'' memory.
\footnote{In WebAssembly-speak, what we refer to as ``flat'' memory, i.e.,
  \mbox{memory that's a sequence of bytes, is called ``linear'' memory}.}
This leads to an important novelty in the type system, which must keep track of
the size of memory slots and disallow strong updates that attempt to store a
larger value in that slot. Hence, capability types in \rwasm track the size
of the memory slot originally allocated and type variables $\alpha$ must be
annotated with a size bound that indicates the maximum size of the type
that $\alpha$ can be instantiated with.

\rwasm is rich enough to serve as a typed compilation target for both
typed garbage-collected languages and languages with an
ownership-based type system and manually managed memory.  We
demonstrate this by implementing compilers from core ML and \lthree to
\rwasm.
\rwasm can be compiled to regular Wasm, providing a pathway to realistic
use in many environments.
We have formalized \rwasm in Coq and we have proved type safety
(progress and preservation) for the type system.

\paragraph{Situating \rwasm.} Before we dive into the details of \rwasm, we
would like to make clear where is sits relative to two other pieces of related
work. The first is the Wasm Component Model, though we've already discussed the
salient difference above, namely ``shared-nothing'' vs. ``fine-grained
shared-memory'' interoperability. But there are additional similarities and
differences. The Component Model is intended to be a part of the WebAssembly
ecosystem but not part of the core Wasm spec and the same is true of
RichWasm. The Component Model provides a means to organize 
Wasm modules into components and instrument core Wasm modules so they can take
advantage of higher-level types when communicating across components, while
RichWasm supports higher-level types and then compiles (or lowers) them to
Wasm. On the other hand, while the Component Model implementation employs
dynamic techniques to assure safety when it comes to features like resources and
handles, RichWasm uses static enforcement.

The second is recent work on safe FFIs by Patterson et al.~\cite{patterson22}
who proposed a framework for design and verification of safe FFIs. Their main
insight is that language designers should build a model of source-level types as
sets of 
target-level terms. Then for all conversions their FFI permits, from a type in
one language say $\tau_A$ to a type in the other say $\tau_B$, the target-level
glue code they write to implement that conversion can be shown to be sound if: 
given target code that behaves like $\tau_A$, the conversion produces target
code that behaves like $\tau_B$.  This is a perfectly reasonable recipe for
designing and verifying safe FFIs between languages that compile directly to
Wasm. However, we would argue that it is a rather heavyweight recipe, one that
requires FFI designers to know how to construct semantic models that would then
guide their thinking. With \rwasm, we would like to provide support for compiler
writers who don't know how to build semantic models and aren't interested in
doing formal verification of FFIs. To that end, we've developed a richly typed IR
capable of detecting unsafe interoperability via type checking of compiled
code. To take advantage of this, compiler writers must implement type-preserving
compilers to \rwasm, a much easier task than defining a target-level model of
source-language types. Whenever the language designer wants to support
additional FFI functionality, they simply have to extend their compiler and see
if any additional conversions they allow result in type-checking errors at the
\rwasm level.

\paragraph{Contributions.} Our central contribution is \rwasm, a typed
IL built on top of WebAssembly that is designed to serve as a
useful platform for safe FFI design.  
\begin{itemize}
\item \rwasm supports an advanced substructural type system with capabilities
  and size tracking that enables static assurance of safe, fine-grained shared
  memory interoperability in a language with a low-level memory model (i.e.,
  ``flat'' memory). The type system allows precisely tracking memory
  ownership and sharing, and avoids memory safety violations even when sharing
  memory across languages with garbage collection and manual memory management
  (\S\ref{sec:overview} and \S\ref{sec:semantics}).
\item We have formalized \rwasm in Coq and proved type safety.
\item We have compilers from core ML and \lthree to
  \rwasm and a simple FFI between them that allows us to compile interoperating programs to \rwasm (\S \ref{sec:compilers}). 
\item We have a compiler from \rwasm to WebAssembly that allows us to run
  \rwasm programs in all hosts of WebAssembly (\S\ref{sec:lowering}).
\end{itemize}

%%% Local Variables:
%%% mode: latex
%%% TeX-master: "main"
%%% End:

%% file: overview.tex
  \negspace
\section{\rwasm overview}
\label{sec:overview}

% Draw similarities with GC proposal
% Explain high-level things that RichWasm is tracking and compare to Wasm
% Call memories owned/unique and shared?
% maybe draw memory layout example
In this section, we will give an overview of the \rwasm language,
presenting its types and syntax. Then we will provide an example of
interoperation of \lthree and ML show how \rwasm can statically detect
safety violations.
First, let's consider a small example to get a sense for the sort of errors
\rwasm can catch. 
In Fig. \ref{fig:simple_example_1} we have a GC'd program which provides two functions:
an identity function on integer references, which stashes a copy of the reference, and a
function which can return the stashed copy.
Our manually managed program first creates a reference, passes it to the $\tt{stash}$
function, and frees the returned reference.
Next, the linear program retrieves the stashed reference and attempts to free
that, resulting in a double free.
If compiled naively, \rwasm's type system will first complain that $\tt{stash}$
requires an unrestricted reference, while it is called with a linear reference.
If $\tt{stash}$ were compiled to take a linear reference, \rwasm would not admit
the function since it duplicates a linear value.

\begin{figure}[htb]
\begin{subfigure}[b]{0.49\textwidth}
\begin{lstlisting}[language=ML,basicstyle=\ttfamily\scriptsize]
let c = ref (ref 0) in
fun stash (r : Ref Int) = c := r; r in
fun get_stashed (() : Unit) = !c
\end{lstlisting}
%% \negspace
\caption{garbage collected program}
\end{subfigure}
\begin{subfigure}[b]{0.49\textwidth}
  \vfill
\begin{lstlisting}[language=ML,basicstyle=\ttfamily\scriptsize,escapeinside={(-}{-)},mathescape]
free (ml.stash (new 42));
free (ml.get_stashed ()) (* CRASH *)
\end{lstlisting}
%% \negspace
\caption{manually managed program}
\end{subfigure}
%% \vspace{3pt}
%% \negspace
\caption{Unsafe interoperability}\label{fig:simple_example_1}
\end{figure}

\paragraph{\rwasm and Wasm.} While \rwasm is coherent
without comparison to Wasm, readers who are familiar with Wasm will note the
parallel structure of the two languages.
Existing constructs from Wasm are extended to support \rwasm's new types, while
continuing to fulfill their original purpose.
For instance, Wasm has "local variables": a location which lives for the duration
of a function call and can store one numeric type. \rwasm has an analagous concept.
However, \rwasm also has strong tools for reasoning about sizes and linearity. 
This allows us to use locals even more effectively, strongly updating them
while guaranteeing that there will be space for any value we
store and that we only duplicate or ignore values which are not linear.

The largest departure from Wasm is in \rwasm's treatment of memory. Because we
want to support strong memory invariants, \rwasm supports a series of structured
memory types, rather than the raw sequence of numeric types present in Wasm.
Wasm's unfettered access to the memory makes it impossible to
guarantee any invariants about one's memory layout when linking with
other code. We no longer support arbitrary memory operations, but
thanks to the same size and linearity reasoning tools that give us
greater control over locals, we still have control over memory layouts
and sharing, without the burden of losing invariants when linking with
other code.

  \negspace
\subsection{Syntax}
In Fig. \ref{fig:terms} we give a full account of \rwasm types and
programs. The highlighted constructs are new in \rwasm and not present
in Wasm.  We start explaining the syntax of the language from the
top-level structures.

\begin{figure}[htb]
\include{terms.tex}
\vspace{-2ex}
\caption{\rwasm abstract syntax}\label{fig:terms}
\vspace{-0.5ex}
\end{figure}

\paragraph{Modules.} As in Wasm, the top-level unit of \rwasm program is a \emph{module}.
A module consists of code in the form of a list of \emph{function}s and data in the
form of a list of \emph{global} declarations that can be mutable or immutable.
Each module also has a \emph{table} that, as in WebAssembly, stores references to
functions which facilitate indirect calls.
Functions, globals, and tables can be exported, making them visible to
other modules for import.

%%% In this section explain some basic concepts before going into the formal description
%%% of the language:
%% - capabilities
%% - memory model: two memories, high-level heap types, garbage collection
%% - sizes
%% - strong updates
%% 
%% Might need to explain some basics about Wasm execution model
%%
%%
%% TODO XXX explain restriction of capabilities in the heap
%%
%%
%% TODO highlight new stuff in the syntax
%%
\paragraph{Functions.} Functions are sequences of \rwasm instructions, taking as input
a sequence of values and returning a sequence of values.
The function type, $\Wfty$, can be polymorphic over various entities
of the type system: memory locations, sizes, qualifiers (which
describe the linearity of a value), and types.
Unlike Wasm, local variables  are not tied to a particular type, so instead they
are defined by their slot size ($\Wsz$) and initialized with an unrestricted
unit value.
Locals may take on linear types during evaluation, and upon use will
be returned to the unrestricted unit type to avoid duplication of a
linear value.

%% \michael{I got rid of the text below and replaced it with a description which
%% relied on less understanding of the rest of the type system:

%% Similar to a struct field, the type of local
%% variable slot can change during the execution of a function. Since a
%% function can consume a linear local value, the value of the should be
%% replaced with a new value, which is of type unit by default.
%% }

\paragraph{Value types.} The types of values consist of a \emph{pretype}
that is annotated with a qualifier. Concrete qualifiers can be linear
($\Wlin$) or unrestricted ($\Wunr$) and indicate whether a value must
be treated linearly. Qualifiers can also be abstract variables bound in function
quantification.
Qualifiers are ordered with $\Wunr \Wqualless \Wlin$. This ordering
helps us put restrictions on abstract types bound in polymorphic,
existential, or recursive types.

Simple pretypes include unit, numeric types (32 or 64 bit signed and
unsigned integers and floats), and tuples $\WPgroup{\many{\Wtau}}$.
Next we have references, pointers, and capabilities. Pointers are used
to access a memory location and capabilities provide ownership to a
memory location. References represent the pair of a capability and a
pointer. Capabilities are a type-level reasoning tool and are erased
in \rwasm programs compiled to Wasm. The ability to split a reference
into a capability and a pointer allows a program to separate its static notion of ownership
from its runtime data layout. If pointers to a location are stashed in
separate parts of a large data strucutre, ownership (in the form of a
capability) can be stored with one of them and then temporarily borrowed
by the other location by moving the capability. Since capabilities will
be erased upon compilation to Wasm, this transfer of ownership
satisifies the type system without incurring any runtime cost.
The type $\WPref{\Wrw}{\Wloc}{\Whty}$ is a reference to a location
$\Wloc$ that contains a \emph{heap type} $\Whty$ and provides read or
read-write priviledge ($\Wrw$) to it. A capability carries similar
information, while a pointer is annotated only with the location that
it points to.
The type $\WPown{\Wloc}$ is an \emph{ownership token} that represents
write ownership of a location. A read-write capability can be temporarily
split into a read-only capability and an ownership token and later
recombined.

In order to represent recursive data structures, \rwasm has
isorecursive types. The type $\WPrec{\Wqual}{\Wptyvar}{\Wtau}$ is
recursive over $\Wptyvar$. The constraint $q \Wqualless$ asserts that
this recursive type will only be unfolded into locations with qualifiers
greater than or equal to the qualifier $q$. The need for such a constraint
is discussed further below.
 
To facilitate location abstraction, which is necessary to statically
represent references to memory locations, \rwasm has existential types
over locations $\WPexists{\Wlocvar}{\Wtau}$. For instance,
$\WPexists{\Wlocvar}{\WPref{\Wrwrw}{\Wlocvar}}{\Whty}$ represents a
read-write reference to a statically unknown location with heap type
$\Whty$.
 
Wasm allows "indirect" function calls, using a runtime value to lookup
a function in a table. We use the type $\WPcoderef{\Wfty}$ to
represent a code pointer to some function in that table with type
$\Wfty$.

Lastly, a pretype can be a pretype variable $\Wptyvar$, referring to
a universal, existential, or recursive binding site.

\paragraph{Heap types and memory model.} The memory model of \rwasm consists
of two global flat memories: the linear memory and the unrestricted
memory. The linear memory is manually managed and references to it
must be treated linearly. The unrestricted memory is garbage collected
and stores ML-like references.
Unlike Wasm, where memories are essentially byte arrays, in \rwasm
memories store high-level structured data. Heap types can be variants,
structs, arrays, and existential packages abstracting over types.
A variant type $\WHvariant{\many{\Wtau}}$ describes a heap value
that can contain more than one kind of value, where $\many{\Wtau}$
is a list containing the type for each case of the variant.
An array type $\WHarray{\Wtau}$ is a variable-length array containing
values of type $\Wtau$.
A struct type $\WHstruct{\many{\Wtau,\ \Wsz}}$ is a record type
with the list $\many{\Wtau,\Wsz}$ indicating the type $\Wtau$ and
\emph{the size} $\Wsz$ of each field.
Keeping the size of each field is necessary to support \emph{strong updates}.
Strong updates are necessary when consuming linear struct fields, as
the old value must be replaced with a new value in order to prevent duplication.
An existential heap type $\WHstruct{\many{\Wtau,\ \Wsz}}$
abstracts over a pretype $\Wptyvar$ in a type $\Wtau$. The qualifier
$q$ denotes the minimum qualifier that the pretype should have when it
is used inside the type. The size $sz$ denotes an upper bound for the
size of the abstracted type, which is useful when we want to \mbox{do a
strong update with this type (either into the field of a struct or a local).}

Locations are natural numbers that refer either to the unrestricted or
linear memory. Since we have location polymorphism, locations can also be
variables.

\paragraph{Function types and polymorphism.} Function types are arrow types $\many{\Wtau_1} \rightarrow \many{\Wtau_2}$
that indicate that a function consumes values with types
$\many{\Wtau_1}$ from the stack, and leaves values with types
$\many{\Wtau_2}$ on the stack after its execution.
To facilitate polymorphism, function types can quantify over four
different kinds: locations, sizes, qualifiers, and
pretypes. Quantification over sizes, qualifiers, and pretypes can
impose constraints over the abstracted kind.

Location quantification is most straightforward: function declarations
should be polymorphic over locations since concrete locations are only
available at runtime.

Functions can also be polymorphic over sizes. For instance, projecting
a struct field is an operation that should be agnostic to the size of the
field, therefore the size must be abstracted.
Quantification over sizes can also be subject to constraints. In
particular, the size variable $\Wszvar$ can be upper-bounded and
lower-bounded by some sizes, written $\many{\Wsz} \Wszless \Wszvar
\Wszless \many{\Wsz'}$.
Allowing quantification over sizes to be constrained in this way
allows for richer reasoning about location usage. For instance, if a
function takes two arguments of sizes $\Wszvar_1$ and $\Wszvar_2$ and
places a tuple consisting of the two arguments into a local of size
$\Wszvar_3$, it must be known that $ \Wszvar_1 + \Wszvar_2 \Wszless
\Wszvar_3 $.

In order to be able to write functions that operate on both linear and unrestricted data, 
functions can also be polymorphic on qualifiers. 
Qualifier quantification can also be subject to the same sort of
contraint as size quantification, in this case written $\many{\Wqual}
\Wqualless \Wqualvar \Wqualless \many{\Wqual'}$.
To see why qualifier constraints are necessary, consider a function
that takes two arguments and constructs a tuple.
The qualifier of the tuple must be greater than the qualifiers on the
values inside the tuple. If this were not the case, we could
have a non-linear tuple which can be duplicated, containing a linear
value which should not be duplicated.

Lastly, we have pretype polymoprhism, which is the usual parametric
polymorphism we find in ML programs. The type constraints are $\Wqual
\Wqualless \Wptyvar^{(c^?)} \Wtyless \Wsz$ where $\Wqual$ is a lower bound for
the qualifier of the pretype, $\Wsz$ an upper bound for its size, and the presence
of $^{c}$ denotes whether this type can contain capabilities.
Type variables having an upper bound on size is straightforwardly useful to demonstrate
that they will fit into locations.
A qualifier lower bound is, however, a bit less intuitive. Keep in
mind that we are quantifying over \emph{pretypes} and not types, so
any location this pretype variable appears will have a qualifier on it. These
bounds provide two guarantees. Firstly, this pretype variable will
only be substituted into positions with qualifiers greater than or
equal to the bound. Secondly, we can only substitute a pretype in for
such a pretype variable if it would be valid at that qualifier.
To see why this is important, consider an attempt to instantiate the variable in the type
$\Wptyvar^{\Wunr}$ with the tuple pretype $\WPgroup{\WPunit^{\Wlin}}$. This instantiation
would leave us with an unrestricted tuple containing a linear value, a type we just
determined would violate the guarantees of our type system. Having quantifier lower bounds
will guarantee that such instantiations never take place.
Whether or not a type contains capabilities is relevant only for garbage collection and
will be discussed below.

Using polymorphism, we can write the type of a function that projects
a linear field of a singleton struct and replaces it with a value of a
new type, performing a strong update:
\[\forall~\Wlocvar~\Wszvar~\Wptyvar_1~(\Wlin\Wqualless\Wptyvar_2^c\Wtyless\Wszvar),~
\Wptyvar_2^{\Wlin}~(\WPref{\Wrwrw}{\Wlocvar}{\WHstruct{(\Wptyvar_1^{\Wlin},\Wszvar)}})^{\Wlin}
\rightarrow
\Wptyvar_1^{\Wlin}~(\WPref{\Wrwrw}{\Wlocvar}{\WHstruct{(\Wptyvar_2^{\Wlin},\Wszvar)}})^{\Wlin}
\]

The function is polymorphic over the actual location of the struct,
$\Wlocvar$, the size of the struct field, $\Wszvar$, the pretype of
the struct field $\Wptyvar_1$, and the pretype of the new value,
$\Wptyvar_2$, that is going to be put in the struct slot.
The constraints over $\Wptyvar_2$ ensure that it will fit into the slot
of size $\Wszvar$ and can be safely stored in the heap. The qualifier
lower bound ensures that this function can be used on types which must
appear in at least linear positions.

\paragraph{Heap values.}
Heap values are variants $\WHVvariant{i}{v}$, where $i$ is tag of
the variant, structs $\WHVstruct{\many{v}}$, arrays
$\WHVarray{i}{\many{v}}$, where $i$ is the size of the array, and
existential packages $\WHVpack{\Wpty}{v}{\Whty}$, where
$\Wpty$ is the pretype witness, $\Whty$ the heap type of the existential
package, and $v$ the value that is being packed.

\paragraph{Values and instructions.} 
Each \rwasm type has a corresponding value form. For unit, numeric
types, and tuples, these values are straightforward. References and
pointers are annotated with the memory location they point
to. Capabilities and ownership tokens carry no information and are
computationally irrelevant. We have the typical constructor for
isorecursive types ($\WVfold{v}$). Existential location packages
($\WVmempack{\Wloc}{v}$) contain the packed value as well as the
location witness of the package, used to instantiate abstract
locations in instructions once unpacked.
The value form for function references is
$\WVcoderef{i}{j}{\many{\Wk}}$, where $i$ is the index of the module
instance, $j$ is the index of the function in the function table, \mbox{and
$\many{\Wk}$ the concrete instantiation of the polymorphic indices.}

New \rwasm instructions include instructions for folding and
unfolding recursive types, packing and unpacking existential memory
locations, grouping and ungrouping tuples, joining and spliting
capabilties and ownership tokens, joining and spliting capabilities
and pointers to form references, and manipulating heap values (each
type of heap value has its own set of instructions).

Instructions that introduce blocks of instructions, like
$\Wkey{block}$ or $\Wkey{if~then~else}$, are annotated not only with
their type as in Wasm, but also with their \emph{local effects},
$\Wleff$. These describe the effect that a list of instructions may
have on the type of their local variable slots: the slot at
position $i$ gets type $\Wtau$. 

  \negspace
\subsection{Unsafe Interoperability}
We extend the languages in the style of \emph{linking types} \cite{patterson17:snapl}, which allow
languages to maintain their native reasoning principles while interacting with foreign types
near language boundaries.
\ml gets a type which directs the compiler to make a type $\tau$ linear in \rwasm, written
$(\tau)^{\Wlin}$. To store linear types, it also has a new construct $\reftolin$ which creates
a reference which can either be empty or contain a linear value of the given type. Normal
$\tt{Ref}$ operations can be used on these references, but \ml compiles them in such a way that
if they are read from or written to twice, they will fail at runtime, as this would violate
linearity and thus not typecheck\footnote{In the following example, the \ml code's use of its
reference to a linear value is entirely valid and would not crash on purpose. This program would
result in a true memory safety violation if admitted by \rwasm.}.
\lthree, which typically only has capabilities and pointers, gets a new \ml-like $\tt{Ref}$ type.
In order to convert to and from this type at the boundary with an \ml program, it also has two
new constructs $\tt{join}$ and $\tt{split}$.

\begin{figure}[htb]
\begin{subfigure}[b]{0.49\textwidth}
\begin{lstlisting}[language=ML,basicstyle=\ttfamily\scriptsize,mathescape]
let c = $\reftolin$ (Ref Int)$^{\Wlin}$ in
fun stash (r : (Ref Int)$^{\Wlin}$) : (Ref Int)$^{\Wlin}$ =
  c := r; r
in
fun get_stashed (() : Unit) = !c
\end{lstlisting}
%% \negspace
\caption{ML program}
\end{subfigure}
\begin{subfigure}[b]{0.49\textwidth}
  \vfill
\begin{lstlisting}[language=ML,basicstyle=\ttfamily\scriptsize,escapeinside={(-}{-)},mathescape]
import (ml.stash :
  !(($\exists$ $\ell$. Ref $\ell$ !Int 1) $\multimap$ ($\exists$ $\ell$. Ref $\ell$ !Int 1)))
import (ml.get_stashed :
  !(!Unit $\multimap$ ($\exists$ $\ell$. Ref $\ell$ !Int 1)))
free (split (stash (join (new !42 1))));
free (split (get_stashed !())) (* CRASH! *)
\end{lstlisting}
%% \negspace
\caption{\lthree program}
\end{subfigure}
%% \vspace{3pt}
%% \negspace
\caption{Unsafe interoperability}\label{fig:full_example_1}
\end{figure}

Fig. \ref{fig:full_example_1} shows an example akin to that in Fig. \ref{fig:simple_example_1},
but with syntax more accurate to \ml and \lthree.
The key differences are that the programs must use their new extensions to agree on types at
the linking boundary. The problematic function here is still \ml's $\tt{stash}$, and \rwasm will not
allow it to typecheck since it duplicates a linear value. If $\tt{stash}$ were to not return the linear
value (and correspondingly \lthree were to no longer attempt to free that value, since it's not
returned), this program would type check and \lthree's previously illegal attempt to free
would be safe.

%% file: terms.tex
{\scriptsize
\begin{flushleft}
\framebox[1.2\width]{Types} \par
\end{flushleft}
\deftable{
 & i &\in& \mathbb{N} \\
\defline{pretypes}{\Wpty}{\mhl{\WPunit}
        		\bnfalt \Wnp
			\bnfalt \mhl{\WPgroup{\many{\Wtau}}}
                        \bnfalt \mhl{\WPref{\Wrw}{\Wloc}{\Whty}}
                        \bnfalt \mhl{\WPptr{\Wloc}}
                        \bnfalt \mhl{\WPcap{\Wrw}{\Wloc}{\Whty}}
                        \bnfalt
%%                         }
%% \defnewline{
                        \mhl{\WPrec{\Wqual}{\Wptyvar}{\Wtau}}
			\bnfalt \mhl{\WPexists{\Wlocvar}{\Wtau}}
                        \bnfalt \mhl{\WPcoderef{\Wfty}}
			\bnfalt \mhl{\WPown{\Wloc}}
                        \bnfalt \mhl{\Wptyvar}}
\defline{numeric pretypes}{\Wnp}{\Wkey{ui32} \bnfalt \Wkey{ui64} \bnfalt \Wkey{i32} \bnfalt \Wkey{i64} \bnfalt \Wkey{f32} \bnfalt \Wkey{f64}}
\defline{types}{\Wtau}{\Wpty^{\mhl{\Wqual}}}
\defline{qualifiers}{\Wqual}{\mhl{\Wqualvar} \bnfalt \mhl{\Wunr} \bnfalt \mhl{\Wlin}}
\defline{memory privilege}{\Wrw}{\mhl{\Wrwrw} \bnfalt \mhl{\Wrwr}}
\defline{heap types}{\Whty}{\mhl{\WHvariant{\many{\Wtau}}}
                    \bnfalt \mhl{\WHstruct{\many{(\Wtau,\ \Wsz)}}}
                    \bnfalt \mhl{\WHarray{\Wtau}}
		    \bnfalt \mhl{\WHexists{\Wqual}{\Wptyvar}{\Wsz}{\Wtau}}}
\defline{locations}{\Wloc}{\mhl{\Wlocvar} \bnfalt  \mhl{i_{\Wunr}} \bnfalt \mhl{i_{\Wlin}}}
\defline{quantifiers}{\Wk}{\mhl{\Wlocvar }
			    \bnfalt \mhl{\many{\Wsz} \Wszless \Wszvar \Wszless \many{\Wsz}}
			    \bnfalt \mhl{\many{\Wqual} \Wqualless \Wqualvar \Wqualless \many{\Wqual}}
			    \bnfalt \mhl{\Wqual \Wqualless \Wptyvar^{(c^?)} \Wtyless \Wsz}}
\defline{arrow types}{\Waty}{\many{\Wtau_1} \rightarrow \many{\Wtau_2}}
\defline{function types}{\Wfty}{\mhl{\forall \many{\Wk}.}~\many{\Wtau_1} \rightarrow \many{\Wtau_2}}
\defline{sizes}{\Wsz}{\mhl{\Wszvar} \bnfalt \mhl{\Wmath{\Wsz + \Wsz}} \bnfalt \mhl{\Wmath{i}}}
%% \defline{local type effects}{\Wleff}{\many{i \rightharpoonup \Wtau}}
%% \defline{Mutable flag}{\Wmut}{\Wmutt \bnfalt \Wmutf}
%% \defline{global types}{\Wglob}{(\Wmut, \Wpty)}
}
\vspace{0.5em}
\begin{flushleft}
\framebox[1.2\width]{Terms} \par
\end{flushleft}
\deftable{
\defline{heap values}{hv}{
                    \mhl{\WHVvariant{i}{v}}
                    \bnfalt \mhl{\WHVstruct{\many{v}}}
                    \bnfalt \mhl{\WHVarray{i}{\many{v}}}
		    \bnfalt \mhl{\WHVpack{\Wpty}{v}{\Whty}}
}
%% }
%% \deftable{
\\
\defline{values}{v}{\mhl{\WVunit}
    \bnfalt np.\Wkey{const}\ c
    \bnfalt \mhl{(\many{v})}
    \bnfalt \mhl{\WVref{\Wloc}}
    \bnfalt \mhl{\WVptr{\Wloc}}
    \bnfalt \mhl{\WVcap} \bnfalt
    \mhl{\WVfold{v}}
    \bnfalt \mhl{\WVmempack{\Wloc}{v}}
    \bnfalt \mhl{\WVcoderef{i}{j}{\many{z}}}
    \bnfalt \mhl{\WVown}
}
%% }
%% \deftable{
\\
\defline{instructions}{e}{v
  \bnfalt np.unopt_{np} 
  \bnfalt np.binop_{np} 
  \bnfalt np.testop_{np} 
  \bnfalt np.cvtop~np' \bnfalt 
  \Wunreach
  \bnfalt \Wnop
  \bnfalt \Wdrop
  \bnfalt \Wselect
  \bnfalt
}
\defnewline{
  \Wblock{\Waty}{\mhl{\Wleff}}{\many{e}} \bnfalt 
  \Wloop{\Waty}{\many{e}}
  \bnfalt \Wite{\Waty}{\mhl{\Wleff}}{\many{e}}{\many{e}} \bnfalt
  \Wbr{i}
  \bnfalt \Wbrif{i}
  \bnfalt \Wbrtable{\many{i}}{j} \bnfaltend
}
\defnewline{
  \Wgetlocal{i}{\Wqual}
  \bnfalt \Wsetlocal{i}
  \bnfalt \Wteelocal{i} \bnfalt
  \Wgetglobal{i}
  \bnfalt \Wsetglobal{i} \bnfalt
  \mhl{\Wqualify {\Wqual}} \bnfaltend
 }
\defnewline{
  \Wreturn \bnfalt
  \mhl{\Wcoderef{i}}
  \bnfalt \mhl{\Winst{\many{\Wk}}}
  \bnfalt \Wcallindirect
  \bnfalt \Wcall{i}{\mhl{\many{\Wk}}} \bnfaltend
}
\defnewline{
  \mhl{\Wrecfold{\Wpty}}
  \bnfalt \mhl{\Wrecunfold} \bnfaltend
            \mhl{\Wmempack{\Wloc}}
  \bnfalt \mhl{\Wmemunpack{\Waty}{\Wleff}{\Wlocvar}{\many{e}}} \bnfalt

}
\defnewline{
  \mhl{\Wseqgroup{i}{\Wqual}}
  \bnfalt \mhl{\Wsequngroup}
  \bnfalt \mhl{\Wcapsplit}
  \bnfalt \mhl{\Wcapjoin} \bnfaltend
  \mhl{\Wrefdemote}
  \bnfalt \mhl{\Wrefsplit}
  \bnfalt \mhl{\Wrefjoin} \bnfaltend
}
\defnewline{
  \mhl{\Wstructmalloc{\many{\Wsz}}{\Wqual}}
  \bnfalt \mhl{\Wstructfree} \bnfaltend
  \mhl{\Wstructget{i}}
  \bnfalt \mhl{\Wstructset{i}}
  \bnfalt \mhl{\Wstructswap{i}} \bnfaltend 
  }
\defnewline{
  \mhl{\Wvariantmalloc{i}{\many{\tau}}{\Wqual}}
  \bnfalt \mhl{\Wvariantcase{\Wqual}{\Whty}{\Waty}{\Wleff}{\many{(\many{e})}}} \bnfaltend
  }
\defnewline{                        
  \mhl{\Warraymalloc{\Wqual}}
  \bnfalt \mhl{\Warrayget}
  \bnfalt \mhl{\Warrayset}
  \bnfalt \mhl{\Warrayfree} \bnfaltend
  }
\defnewline{                        
  \mhl{\Wexistpack{\Wpty}{\Whty}{\Wqual}}
  \bnfalt \mhl{\Wexistunpack{\Wqual}{\Whty}{\Waty}{\Wleff}{\Wpty}{\many{e}}} \bnfaltend
}
}
\hspace{-2em}
\begin{minipage}{0.54\textwidth}
%% \hspace{-2em}
\deftable{
\defline{}{unop_{\Wkey{i}N}}{
  \Wkey{clz}
  \bnfalt
  \Wkey{cnt}
  \bnfalt
  \Wkey{popcnt}
}
\defline{}{binop_{\Wkey{i}N}}{
  \Wkey{add}
  \bnfalt
  \Wkey{sub}
  \bnfalt
  \Wkey{mul}
  \bnfalt
  \Wkey{div}~\Wsign \bnfalt
  \Wkey{rem}~\Wsign \bnfaltend
  }
\defnewline{
  \Wkey{and} \bnfalt
  \Wkey{or}
  \bnfalt
  \Wkey{xor} \bnfalt 
  \Wkey{shl} \bnfalt
  \Wkey{shr}
  \bnfalt
  \Wkey{rotl}
  \bnfalt
  \Wkey{rotr}
}
\defline{}{unop_{\Wkey{f}N}}{
  \Wkey{abs}
  \bnfalt
  \Wkey{neq}
  \bnfalt
  \Wkey{sqrt}
  \bnfalt
  \Wkey{ceil} \bnfalt 
  \Wkey{floor}
  \bnfalt
  \Wkey{trunc}
  \bnfalt
  \Wkey{nearest}
}
\defline{}{binop_{\Wkey{f}N}}{
  \Wkey{add}
  \bnfalt
  \Wkey{sub}
  \bnfalt
  \Wkey{mul}
  \bnfalt
  \Wkey{div} \bnfalt 
  \Wkey{min}
  \bnfalt
  \Wkey{max}
  \bnfalt
  \Wkey{copysign}
}
}
\end{minipage}
\begin{minipage}{0.45\textwidth}
\deftable{
\defline{}{testop_{\Wkey{i}N}}{\STEITeqz}
\defline{}{relop_{\Wkey{i}N}}{
  \STEIReq
  \bnfalt \STEIRne
  \bnfalt \STEIRlt{\Wsign} \bnfalt
\STEIRgt{\Wsign}
  \bnfalt \STEIRle{\Wsign}
  \bnfalt \STEIRge{\Wsign}}
\defline{}{relop_{\Wkey{f}N}}{
  \STEFReq
  \bnfalt \STEFRne
  \bnfalt \STEFRlt
  \bnfalt
\STEFRgt
  \bnfalt \STEFRle
  \bnfalt \STEFRge}
\defline{}{cvtop}{
  \Wkey{convert}
  \bnfalt
  \Wkey{reinterpret}
}
}
\end{minipage}

%% \defline{administrative instructions}{e}{\dots
%%   \bnfalt \Wtrap
%%   \bnfalt \Wcallcl{\Wcl}{\many{\Wk}}                                         
%% }
%% \defnewline{
%%   \Wlabel{i}{\Waty}{\Wleff}
%% 	  {\many{e_1}}
%%           {\many{e_2}}
%%   \bnfalt \Wlocal{i}
%% 	  {j}
%%           {\many{(v, \Wsz)}}
%%           {\many{e}}
%% }
%% \defnewline{
%%   \Wmalloc{\Wsz}{hv}{\Wqual}
%%   \bnfalt \Wfree
%% }
\vspace{0.5em}
\begin{flushleft}
\framebox[1.1\width]{Top-level declarations} \par
\end{flushleft}
\vspace{-0.5em}

\begin{minipage}{0.6\textwidth}
\begin{flushleft}
\deftable{
\defline{functions}{f}{\many{ex}~\Wkey{function}~\Wfty~\Wkey{local}~\mhl{\many{\Wsz}}~\many{e}
\bnfalt \many{ex}~\Wkey{function}~im}

\defline{globals}{glob}{\many{ex}~\Wkey{glob}~\Wkey{mut^{?}}~\Wpty~\many{i}
                        \bnfalt \many{ex}~\Wkey{glob}~im}
\defline{table}{tab}{\many{ex}~\Wkey{table}~\many{i}
                     \bnfalt \many{ex}~\Wkey{table}~im}
}
\end{flushleft}
\end{minipage}
\hfill
\begin{minipage}{0.39\textwidth}

\deftable{
\defline{exports}{ex}{\Wkey{export}~\text{``}name\text{''}}
\defline{imports}{im}{\Wkey{import}~\text{``}name\text{''}}
\defline{modules}{m}{\Wkey{module}~\many{f}~\many{glob}~tab}
}
\end{minipage}
}

%% file: dynamics.tex
  \negspace
\section{\rwasm Dynamic Semantics}\label{sec:semantics}
Execution in \rwasm terms closely follows Wasm and it is defined as a
reduction relation. The relation, written
$\Wredi{s}{\many{v}}{\many{\Wsz}}{\many{e}}{s'}{\many{v'}}{\many{e'}}$,
represents a reduction step in module $j$ from one program configuration
$s;\many{v};\many{sz};\many{e}$ to another, where $s$ is the store, $\many{v}$
the local values and $\many{sz}$ the sizes of their slots, and
$\many{e}$ the instructions to be evaluated. Fig. \ref{fig:red} shows the definition of runtime
objects as well as important rules of the reduction relation.
We elide rules that are identical to Wasm, and we focus on the
reduction rules of new constructs.

\begin{figure}[p]
  \include{exec.tex}
  \caption{\rwasm dynamic semantics}\label{fig:red}
\end{figure}

\paragraph{Store and module instances.} A store represents the execution 
state. It holds the list of \emph{module instances}, which is the
dynamic representation of static modules and the \emph{global
memory}. A module instance consists of a dynamic function table that
holds a list of \emph{closures} which, as in Wasm, is used for direct
calls, a list of global values, and the dynamic table which is a list
of closures that is used for indirect calls. A closure, in the Wasm
sense, is a dynamic instance of a function that consists of the
code and a pointer to the module that provides the function's
environment.
The global memory has two components: the \emph{linear} memory and the
\emph{unrestricted} memory. Both memories are maps from locations to
high-level heap values.

%% These should be explained earlier
%% %
%% The shared memory is garbage collected and
%% references to it can be freely duplicated and shared.
%% %
%% The \emph{owned} memory is manually managed and references to it are
%% linear and uniquely owned at any given point during a program's
%% execution.
%% %
%% The memory address space is shared across all \rwasm modules, as
%% opposed to Wasm, where modules have their own memory address space and
%% can be shared with another module only by directly importing and
%% exporting it. Because \rwasm enforces strict control over memory access,
%% no such separation is necessary.

\paragraph{Administrative instructions.} As in Wasm, some reduction rules
generate instructions that are not part of the syntax of source
programs and represent administrative operations. A $\Wtrap$
instruction signals an execution trap. We add two administrative instructions: 
$\Wmalloc{\Wsz}{hv}{\Wqual}$ to allocate memory of size $\Wsz$ holding a value $hv$
in the memory $q$, and $\Wfree$ to deallocate parts of the \emph{linear} memory, given
a reference. Allocation and deallocation of aggregate types reduces to these
administrative instructions.
The administrative instruction $\Wcallcl{\Wcl}{\many{z}}$ is similar
to Wasm's but along with the closure $cl$, it is annotated with the
concrete instantiation of the polymorphic quantifiers in the
function's type, $\many{z}$.

\paragraph{Control flow.}
Much like in Wasm, when running \rwasm programs, evaluation occurs
within some number of label instructions, each of which corresponds to
a source block of code (introduced by instructions like
$\Wkey{block}$, $\Wkey{if}$, $\Wkey{loop}$). The break instructions
($\Wkey{br}$, $\Wkey{br\_if}$, $\Wkey{br\_table}$) allow programs to
jump to any of the surrounding N locations by specifying how many
labels to jump over. Nested label instructions are represented with
$\emph{local contexts}$. A local context $\locctx^N[\many{e}]$
represents N nested label instructions. A local context has a hole at
its most deep label instruction, where evaluation can occur.

%% Our type system imposes constraints on which of
%% these jumps are valid. Firstly, when jumping to a label, one must have
%% the correct stack type and local context type. This is enforced by the
%% label context. Secondly, a break must not jump over, and thus drop,
%% any potentially linear values. This is enforced by the linear context.
%% Local contexts represent nested labels, and evaluation proceeds inside
%% the most deeply nested label until either a break instruction explicitly
%% jumps to a surrounding label or the inner instruction can be reduced
%% to a list of values or a trap.
%% \zoe{TODO maybe something about branches?}
%% \michael{I think this is sufficient now for the operational semantics.
%% We'll want to talk more about branching when we discuss the static semantics. }

  \negspace
\subsection{Reduction relation.} 
We give an overview of the most important reduction rules
of \rwasm that are new or different from those in Wasm.
Additional reduction rules can be found in \S 1 of
the appendix. The full set of rules is encoded in our Coq development.
For space reasons, we drop the store and local variables in all the
rules that leave them unchanged. We may still use the store $s$ in
side conditions where it is relevant, without explicitly mentioning it
in the rule.

\paragraph{Garbage collection.}
The reduction relation is equipped with a rule that can be applied at
any point and allows collection of unrestricted locations that are no
longer accessible from the configuration. Therefore, the roots of collection
are the unrestricted locations that appear in reference values in the
instructions, local variables, or the module instances. Any location not
reachable from the roots is collected.

If a reference to linear memory is placed into garbage collected memory,
we say that the garbage collector now \emph{owns} that memory and is
responsible for collecting it if the unrestricted location, and thus the
only reference to the linear location, should be collected.
Freeing of linear memory is a type-directed operation, and when compiling
to Wasm, we can generate finalizer functions that get called when such
references are collected. But what would we do if a capability to linear
memory were in garbage-collected memory? When compiling to Wasm, capabilities
will be erased, which would leave the garbage collector with no way to reference
the linear memory location it owns at runtime. To resolve this, we require
that capabilities always be paired with a pointer in the form of a reference
when placed in memory.

\paragraph{Function calls.} Function calls follow similar
principles with function calls in Wasm. Few differences arise from the
polymorphic types of \rwasm functions and the way polymorphic binders
are instantiated.
Direct function calls are performed with the $\Wcall{j}{\many{z}}$
instruction, where $j$ is the index of the function in the function
table and $\many{z}$ that concrete instantiations of the polymorphic
binders.
The instruction $\Wcall{j}{\many{z}}$ is reduced to the administrative
instruction $\Wcallcl{\mathit{cl}}{\many{z}}$, where $\mathit{cl}$ is
the corresponding closure in the function table.
Indirect function calls are performed with $\Wcallindirect$, whose
argument is a code pointer $\WVcoderef{i}{j}{\many{z}}$ that points to
the $j$th function of the $i$th module instance and $\many{z}$ are the
concrete instantiations of polymorphic indexes.
%
%% The instantiation of polymoprhic indexed of code pointers happens with
%% the $\Winst$ instruction (rule $(4)$).
% TODO  
%% %
%% To construct a code pointer value, $\rwasm$ programs use the
%% $\Wcoderef{j}$ instruction that constructs the code pointer to the
%% $j$th function of the current module instance. 

The administrative call instruction $\Wcallcl{\mathit{cl}}{\many{z}}$,
where $\mathit{cl}$ is a closure containing a module number and function
body, takes as stack arguments the arguments of the function and reduces to
a local frame that performs the necessary substitutions of the
polymorphic indices. For the sizes of local variable slots we use the
metafunction $\sz(\cdot)$ that returns the size of a type.

\paragraph{Heap manipulation.}
Let's consider the rules for manipulating \rwasm's new heap structures.
All heap data structures have have their own allocation
instruction that reduces to the administrative malloc instruction we saw above.
This malloc instruction then reduces to a reference contained in an
existential package which abstracts the location.

Reduction rules for struct's get, set and swap operations are straightforward,
taking a reference from the stack and perhaps the value to put at the
$i$th field, performing any necessary memory updates, and leaving the
reference back on the stack, together with the read value, if any.

Variants can be used to perform case analysis. The case instruction expects on the
stack the reference to the variant and a list of values expected by the branches,
that have types $\manyn{\Wty_1}{n}$. 
If the allocated value is the $i$th case of the variant, then we create
an instruction block using the $i$th element of the list of instruction
blocks $\manyn{(\many{e})}{m}$, written $\manyn{(\many{e})}{m}_{(i)}$.
If the case instruction is anntotated with an unrestricted qualifier, the
reference is returned to the stack for reuse. 
If the annotation is linear, then a $\Wfree$ instruction is generated to consume
the reference. In this case the contents of the memory are
replaced with an empty array, so that the linearity invariants of the
type system are preserved.
Existential types can be packed and unpacked. Packing an existential
type triggers the allocation of a $\Wkey{pack}$ heap value.
Much like variants, the unpacking operation is a block instruction
which can optionally free the underlying memory. The unpack operation
additionally needs to substitute the witness pretype in the list of
instructions to be evaluated.

%% Structs can be allocated, freed, read from, written to, and
%% \emph{swapped}. Reading the $i$th element of the struct happens with
%% the instruction $\Wstructget{i}$ that takes a reference from the stack
%% and leaves it back on the stack together with the read value.
%
%% Setting the $i$th element of the struct happens with the instruction
%% $\Wstructset{i}$. \michael{I took out the discussion of what needs to be
%% unrestricted because it's more of a typing distinction. There is no restriction
%% on qualifier in the reduction rule.}
%% %
%% Reading or writing a linear value from or to a struct cell can happen
%% only with the $\Wstructswap{i}$ instruction that performs both operations
%% at the same time, to prevent a potentially linear value from being present
%% simultaneously both on the stack and in the heap.

%%% Local Variables:
%%% mode: latex
%%% TeX-master: "main"
%%% End:

%% file: exec.tex
{\scriptsize
\begin{minipage}{0.45\textwidth}
\begin{flushleft} 
\deftable{
\defline{store}{s}{ \{ \Wsf{inst}~ \many{inst},  \Wsf{mem}~ mem \} }
\defline{instances}{inst}{\{  \Wsf{func}~ \many{cl}, \Wsf{glob}~\many{v}, \Wsf{tab} ~ \many{sl} \} }
}
\end{flushleft}
\end{minipage}
\begin{minipage}{0.45\textwidth}
\begin{flushleft}
\deftable{
\defline{memory}{mem}{\{  \Wsf{lin}~ i \mapsto hv ,  \Wsf{unr}~ i \mapsto hv \} }
\defline{closure}{cl}{\{  \Wsf{inst}~i, \Wsf{code}~f \}}}
\end{flushleft}
\end{minipage}

\deftable{
\\
\defline{administrative instructions}{e}{\dots
  \bnfalt \Wtrap
  \bnfalt \Wcallcl{\Wcl}{\many{z}}
  \bnfalt
  \Wlabel{i}{\Waty}
	  {\many{e_1}}
          {\many{e_2}}  \bnfalt
}
\defnewline{
  \Wlocal{i}
	  {j}
          {\many{(v, \Wsz)}}
          {\many{e}}
  \bnfalt
  \Wmalloc{\Wsz}{hv}{\Wqual}
  \bnfalt \Wfree
}
\defline{local contexts}{\WL{0}}{\many{v}~[\_]~ \many{e}}
\defline{}{\WL{k+1}}{\many{v}~\Wlabel{i}{\Waty}{\many{e}}{\WL{k}}~\many{e}} 
}

\begin{flushleft}
{\bf Reduction } \hfill $\boxed{\Wredi{s}{\many{v}}{\many{\Wsz}}{\many{e}}{s'}{\many{v'}}{\many{e'}}}$ 
\end{flushleft}

\begin{mathpar}
\infer []
{\Wredi{s}{\many{v}}{\many{sz}}{\many{e}}{s'}{\many{v'}}{\many{e'}}}
{\Wredi{s}{\many{v}}{\many{sz}}{\lapp{\many{e}}}{s'}{\many{v'}}{\lapp{\many{e'}}}}
\hspace{0.5cm}
\infer []
{\Wredi{s}{\many{v}}{\many{\Wsz}}{\many{e}}{s'}{\many{v'}}{\many{e'}}}
{
\Wred{s}{\many{v_0}}{\many{i}}{\Wlocal{n}{i}{\many{(v, \Wsz)}}{\many{e}}}
     {s'}{\many{v_0}}{\Wlocal{n}{i}{\many{(v', \Wsz)}}{\many{e'}}}{j}
} \\
\hspace{0.5cm}
\infer []
{
\mathbf{collect}(\locs{\many{e}}\cup\locs{\many{v}}\cup\locs{inst},mem, mem'), 
}
{
\Wredi
{\{ \Wsf{inst}~ \many{inst},  \Wsf{mem}~ mem\}}
{\many{v}}{\many{\Wsz}}{\many{e}}
{\{ \Wsf{inst}~ \many{inst},  \Wsf{mem}~ mem'\}}
{\many{v}}{\many{e}}
}
\end{mathpar}

\[\begin{array}{r l l}
% call
\redrulei {s;\many{v};\many{sz}; \Wcall{j}{\many{z}}}
{s;\many{v};\Wcallcl{cl}{\many{z}}}
\hfill \text{where } cl = (s_{\mathsf{inst}}(i))_{\mathsf{tab}}(j) \\ 
% call indirect
\redrule {s;\many{v};\many{sz}; \WVcoderef{i}{j}{\many{z}}; \Wcallindirect}
{s;\many{v};\Wcallcl{cl}{\many{z}}} \hfill \text{where } cl = (s_{\mathsf{inst}}(i))_{\mathsf{tab}}(j) \\
% call admin 
\redrule{\manyn{v}{n}~\Wcallcl{\{\Wsf{inst}~i, \Wsf{code}~{\Wkey{function}~\forall \many{\Wk}.~\manyn{\Wtau_1}{n} \rightarrow \manyn{\Wtau_2}{m}~\Wkey{local}~\manyn{\Wsz}{k}~\many{e}} \}}{\many{z}}}
{\Wlocal{m}{i}{locals}{\many{e}\sbst{\many{\Wk}}{\many{z}}}} \hfill \text{where} \\ 
& & \hfill locals = \manyn{(v,{\sz{\Wtau_1\sbst{\many{\Wk}}{\many{z}}}})}{n} \manyn{(\WVunit, {\Wsz}\sbst{\many{\Wk}}{\many{z}})}{k} \\

% alloc
\redrule{\manyn{v}{n}~\Wstructmalloc{\manyn{\mathit{sz}}{n}}{q}}{\Wmalloc{\sz{\manyn{\mathit{sz}}{n}}}{\WHVstruct{\manyn{v}{n}}}{q}} \\
% struct free
\redrule{\Wstructfree}{\Wfree} \\\
% struct get
\redrule{(\WVref{l_{mem}})~\Wstructget{i}}
{s;\many{v};(\WVref{l_{mem}})~v_i}
\hfill \text{where }  \\
& & \hfill {(s_{\Wkey{mem}})_{mem}(l) = \WHVstruct{v_1 \dots v_i \dots}} \\
% struct set
\redrule{s;\many{v};\many{sz};(\WVref{l_{mem}})~v~\Wstructset{i}}
{s';\many{v};\WVref{l_{mem}}}
\hfill  \text{where }  \\
& & \hfill {(s_{\Wkey{mem}})_{mem}(l) = \WHVstruct{v_1 \dots v_i \dots}} \\
& & \hfill {s' = s \text{ with } \Wkey{mem}_{mem}(l) = \WHVstruct{v_1 \dots v_{i-1}~ v \dots}} \\
% struct swap
\redrule{s;\many{v};\many{sz};(\WVref{l_{mem}})~v~\Wstructswap{i}}
{s';\many{v};(\WVref{l_{mem}})~v_i}
\hfill  \text{where }  \\
& & \hfill {(s_{\Wkey{mem}})_{mem}(l)= \WHVstruct{v_1 \dots v_i \dots}} \\
& & \hfill {s' = s \text{ with } \Wkey{mem}_{mem}(l) = \WHVstruct{v_1 \dots v_{i-1}~ v \dots}} \\
%variants
 \redrule{v~\Wvariantmalloc{j}{\many{\Wty}}{q}}{\Wmalloc{(32 + \sz{v})}{\WHVvariant{j}{v}}{q}} \\
% variant case unr
\redrule
{
\WVref{l_{\Wunr}}~\manyn{v}{n}~\Wvariantcase{\Wunr}{\Whty}{\Waty}{\Wleff}
{\manyn{(\many{e})}{m}}}
{(\WVref{l_{\Wunr}})~\manyn{v}{n}~\Wblock{\Waty}{\Wleff}{v'~\manyn{(\many{e})}{m}_{(i)}}} \\
 & & \hfill \text{where } \Whty = \WHvariant{\manyn{\Wtau}{m}} \\
& & \hfill \Waty = \Warrow{\manyn{\Wtau_1}{n}}{\many{\Wtau_2}} \\ & & \hfill {(s_{\Wkey{mem}})_{\Wunr}(l)  = \WHVvariant{i}{v'}} \\
% variant case lin
 \redrule{
  s;\many{v};\many{sz};(\WVref{l_{\Wunr}})~\manyn{v}{n}~\Wvariantcase{\Wlin}{\Whty}{\Waty}{\Wleff}
  {\manyn{(\many{e})}{m}}
  }
  {s';\many{v};(\WVref{l_{\Wlin}})~\Wfree~\manyn{v}{n}~\Wblock{\Waty}{\Wleff}{v'~\manyn{(\many{e})}{m}_{(i)}}}  \\
 & & \hfill \text{where } \Whty = \WHvariant{\manyn{\Wtau}{m}} \\
 & & \hfill \Waty = \Warrow{\manyn{\Wtau_1}{n}}{\many{\Wtau_2}} \\
 & & \hfill  (s_{\Wkey{mem}})_{\Wlin}(l)  = \WHVvariant{i}{v'} \\ 
 & & \hfill s' = s \text{ with } \Wkey{mem}_{\Wlin}(l) = \WHVarray{0}{\epsilon} \\
% arrays 
 \redrule{v~\Wkey{ui32.const}~n~\Warraymalloc{q}}{\Wmalloc{(j\times\sz{v})}{\WHVarray{j}{\manyn{v}{j}}}{q}} \\
% array get
 \redrule{(\WVref{l_{mem}})~(np.\Wkey{const}\ j)~\Warrayget}
{(\WVref{l_{mem}})~v_j} \\
 & & \hfill  \text{where }{(s_{\Wkey{mem}})_{mem}(l)  = \WHVarray{i}{(v_0 \dots v_j \dots)}} \\
% array get trap
 \redrule{(\WVref{l_{mem}})~(np.\Wkey{const}\ j)~\Warrayget}
{\Wtrap} \\
 & & \hfill \text{where }  {(s_{\Wkey{mem}})_{mem}(l) = \WHVarray{i}{\many{v}}} \\
 & & \hfill {j\geq i\text{ or }j < 0} \\
% array set
 \redrule{s;\many{v};\many{sz};(\WVref{l_{mem}})~(np.\Wkey{const}\ j)~v~\Warrayset}
{s';\many{v};\WVref{l_{mem}}} \\
 & & \hfill \text{where } {(s_{\Wkey{mem}})_{mem}(l) = \WHVarray{i}{v_0\dots v_{j-1} v_j\dots}} \\
 & & \hfill {s' = s \text{ with } \Wkey{mem}_{mem}(l) = \WHVarray{i}{v_0\dots v_{j-1} v\dots}} \\
% array set trap
 \redrule{(\WVref{l_{mem}})~(np.\Wkey{const}\ j)~\Warrayset}
{\Wtrap} \\
 & & \hfill \text{where }  (s_{\Wkey{mem}})_{mem}(l)= {\WHVarray{i}{\many{v}} } \\
 & & \hfill {j\geq i\text{ or }j < 0} \\
% exists 
 \redrule{v~\Wexistpack{\Wpty}{\Whty}{q}}{\Wmalloc{(64+\sz{v})}{\WHVpack{\Wpty}{v}{\Whty}}{q}}\\
% unpack unr
 \redrule{(\WVref{l_{\Wunr}})~\manyn{v}{n}~\Wexistunpack{\Wunr}{\Whty}{\Waty}{\Wleff}{\Wptyvar}{\many{e}}}
{(\WVref{l_{\Wunr}})~\manyn{v}{n}~
  \Wblock{\Waty}{\Wleff}{v'~{\many{e}}\sbst{\Wptyvar}{\Wpty}}} \\
 & & \hfill{\text{where } (s_{\Wkey{mem}})_{\Wunr}(l) = \WHVpack{\Wpty}{v'}{\Whty}} \\
 & & \hfill \Waty = {\Warrow{\manyn{\Wtau_1}{n}}{\many{\Wtau_2}}} \\
% unpack lin
 \redrule{s;\many{v};\many{sz};(\WVref{l_{\Wlin}})~\manyn{v}{n}~\Wexistunpack{\Wlin}{\Whty}{\Waty}{\Wleff}{\Wptyvar}{\many{e}}}
 {s';\many{v};(\WVref{l_{\Wlin}})~\Wfree
  ~\manyn{v}{n}~\Wblock{\Waty}{\Wleff}{v'~{\many{e}}\sbst{\Wptyvar}{\Wpty}}} \\
 & & \hfill{\text{where } (s_{\Wkey{mem}})_{\Wlin}(l)\WHVpack{\Wpty}{v'}{\Whty} } \\
 & & \hfill {s' = s \text{ with } \Wkey{mem}_{\Wlin}(l) = \WHVarray{0}{\epsilon}} \\
 & & \hfill \Waty = \Warrow{\manyn{\Wtau_1}{n}}{\many{\Wtau_2}} \\
% free
 \redrule{s;\many{v};\many{sz};(\WVref{l_{\Wlin}})~\Wfree}{s';\many{v};\epsilon}
\hfill  \text{~where }  \\
 & & \hfill {s' = s \text{ with } l\notin\text{dom}(\Wkey{mem}_{\Wlin})} \\
% malloc
 \redrule{s;\many{v};\many{sz};\Wmalloc{\Wsz}{\Whv}{q}}{s';\many{v};\WVmempack{\Wloc_{q}}{(\WVref{\Wloc_{q}})}} \\
 & & \hfill \text{where } {s' = s \text{ with } \Wkey{mem}_{q}(\Wloc) = \Whv} \\
% malloc trap
\end{array} 
\]}
\vspace{-3em}

%% file: statics.tex
  \negspace
\section{\rwasm Type System}
\label{sec:statics}
We now give a technical account of the type system as well as its safety properties.

\paragraph{Typing environments.}
Typing environments in \rwasm (fig. \ref{fig:tenv}) are similar to
Wasm, but they also keep track of the new kind variables (types,
sizes, locations, qualifiers) and their constraints.

The \emph{local} environment keeps track of the type and size of the
local variables of a program configuration. It is a list of a type and
size where the $i$th element corresponds to the size and the type of
local variable $i$.

The \emph{function environment} is used to give a function type to a
list of an expressions. It has 7 components. As in Wasm, the
$\mathsf{label}$ component keeps track of the return type of all the
available jump locations (i.e., nested labels). In \rwasm, it additionally
tracks the resulting local environment, as all jumps to a location must
have the same view of the types of locals.
The $\mathsf{return}$ component keeps track of the return type of the
execution of the current block of instructions.
$\mathsf{qual}$, $\mathsf{size}$, $\mathsf{type}$ are partial maps from 
the variables in scope to the constraints placed on them, as explained
in \S \ref{sec:overview}.
The $\mathsf{location}$ component keeps track of declared location
variables.
Lastly, the $\mathsf{linear}$ environment contains a list of
qualifiers representing the greatest lower bound of the qualifiers of
the values on the stack between two jump locations. Jumping from the
current evaluation context to an outer label will drop the contents of
that label, including any potentially linear values. The linear
environment can be used to verify that all values dropped when
performing a jumping to a given label are unrestricted.

To update an component of a typing environment we use similar notation
as the WebAssembly paper. For instance, we write $F, \mathsf{linear}~
\Wunr :: F_{\mathsf{linear}}$ to update the linear environment of $F$
by inserting the qualifier $\Wunr$ to the top of the list.

\begin{figure}[htb]
  {\scriptsize
 \deftable{
  \defline{Local Environment}{\locctx}{ \many{(\Wtau, \Wsz)}}
  \defline{Function Environment}{\funcctx}{\{\
    \mathsf{label}\ \many{(\many{\Wty}, \locctx)},\
    \mathsf{return}\ (\many{\Wty})^{?},\ 
    \mathsf{qual}\ \Wqualvar \rightharpoonup (\many{\Wqual}, \many{\Wqual}), \ 
    \mathsf{size}\ \Wszvar \rightharpoonup (\many{\Wsz}, \many{\Wsz}), \ \\&&& \ \ \ 
    \mathsf{type}\ \Wptyvar \rightharpoonup (\Wsz, \Wqual, hc),\ %%TODO fix heapable constant
    \mathsf{location}\ \many{\Wloc},\
    \mathsf{linear}\ \many{\Wqual} \ \}
  }
  \defline{Module Environment}{\modctx}{
    \{\ \mathsf{func}\ \many{\Waty},\ 
    \mathsf{global}\ \many{\Wglob},\ 
    \mathsf{table}\ \many{\Waty} \ \}
  }
  \defline{Store Typing}{\storetyp}
          {\{\ \mathsf{inst}\ \many{\modctx},\
            \mathsf{unr}\ \Wloc \rightharpoonup \Whty,\
            \mathsf{lin}\ \Wloc \rightharpoonup \Whty
            \ \}
          }
}}
  \negspace
\caption{Typing environments.}\label{fig:tenv}
\end{figure}

The module environment keeps track of the declared functions and globals in
the current module.
Finally, we have the store typing that keeps track of the list of
module instances ($\mathsf{inst}$) and the typing of the linear and unrestricted memories.
The memory typing is a partial map from a memory locations to a pair
of the stored heap type ($\Whty$) and the size of the slot.
The linear memory typing is split across
the typing of the subexpressions in the premises of a rule to ensure
that no linear resource is used twice in the program.
We write $S= S_1 \uplus S_2 $ to denote that the linear memory typing
of $S_{ \mathsf{lin}}$ is the \emph{disjoint} union of
$(S_1){_\mathsf{lin}}$ and $(S_2){_\mathsf{lin}}$, whereas the other
components are exactly the same in all three store typings.
When typing a base value or instruction that has no linear memory
locations, we require that the \mbox{linear store typing is empty, ie., no
linear resource is ever dropped.}

\paragraph{Value Typing.} The value typing judgment, written
$\valtype{S}{F}{v}{\Wty}$, asserts that a value $v$ has type $\Wty$
in a store typing $S$ and function environment $F$. 
Selected value typing rules are shown in Fig.
\ref{fig:valtyp}. Numeric constants have the corresponding numeric
type and can have any qualifier.
Tuples have a tuple type consisting of the types of individual values.
The top-level qualifier must be an upper bound for each individual
qualifier $q_i$ of any type $p_i^{q_i}$ inside the list $\many{\Wty}$.
Pointers have type $\WPptr^{\Wqual}$, which is independent of the heap
type of the location, and do not consume a location from the memory
typing as they do not represent memory ownership.
Typing of capabilities and references are similar. We only explain
references.
In the unrestricted case, a reference $\WVref{\Wlocunr}$ has type
$(\WPref{\Wrw}{\Wlocunr}{\Whty})^{\Wqual}$, where $\Whty$ is the type
of $\Wlocunr$ in the unrestricted component of the heap. The qualifier
$q$ must be provably unrestricted in the constraints of
$F_{\mathsf{qual}}$, written $\Wqual \Wquallessctx \Wunr$. The linear
component of the memory typing must be empty.
The linear case is similar. The type $\Whty$ is the type of $\Wloclin$
in the linear component of the heap, which must be a singleton. The
qualifier must be linear.
The value $\WVfold{v}$ has the recursive type
$\WPrec{\Wptyvar}{\Wqual}{\Wpty^\Wqual})^{\Wqual'}$, if the value $v$
has type
$\Wpty\subst{\Wptyvar}{\WPrec{\Wptyvar}{\Wqual}{\Wpty^\Wqual}})^\Wqual$
where we have substituted $\Wptyvar$ with
$\WPrec{\Wptyvar}{\Wqual}{\Wpty^\Wqual}$, and the qualifier $\Wqual$
is upper bounded by $\Wqual'$. 
Next we have existential packages $\WVmempack{\Wloc}{v}$ that have an
existential type $(\WPexists{\Wlocvar}{\Wpty^\Wqual})^{\Wqual'}$ if
the type of value $v$ is $(\Wpty\subst{\Wlocvar}{\Wloc})^{\Wqual'}$
and qualifier $\Wqual$ is upper bounded by $\Wqual'$.
Lastly, we have typing for code references
$\WVcoderef{i}{j}{\many{\Wk'}}$: if $M$ is the $i$th module
instance and $\forall \many{\Wk}. \Waty $ the type of the $j$th
function in the table of $M$, then the code reference has type
$(\Waty\subst{\many{\Wk}}{\many{\Wk'}})^\Wqual$ where
quantifier variables $\Wk$ have been substituted with the indices
$\Wk'$.

\begin{figure}[htb]
  {\scriptsize
    \begin{flushleft}
    \fbox{$\valtype{\storetyp}{\funcctx}{v}{\Wty}$}
  \end{flushleft}
\begin{mathpar}
         \infer []
       {S = \biguplus_{i\leq n}S_i \\
         \forall~v_i\in\manyn{v}{n}~\Wty_i\in \manyn{\Wty}{n}, ~\valtype{\storetyp_i}{\funcctx}{v_i}{\Wty_i} \\
         \forall~\Wpty_i^{\Wqual_i} \in\manyn{\Wty}{n}, ~ \Wqual_i \Wquallessctx \Wqual }
       { \valtype{\storetyp}{\funcctx}{\manyn{v}{n}}{(\manyn{\Wty}{n})^{\Wqual}}}
  \and \infer []
       {\linempty}
       {\valtype{\storetyp}{\funcctx}{\WVptr{\Wloc}}{\WPptr^\Wqual}}
  % \and \infer []
  %      {\linempty \\
  %       \gethtunr{\Wloc}{\Whty} \\
  %       \Wqual \Wquallessctx \Wunr
  %      }
  %      {\valtype{\storetyp}{\funcctx}{\WVref{\Wloc}}{\Wqual~\WPref{\Wrw}{\Wloc}{\Whty}}}
  \and \infer []
       {\linempty \\
        \gethtunr{\Wloc}{\Whty} \\
        \Wqual \Wquallessctx \Wunr
       }
       {\valtype{\storetyp}{\funcctx}{\WVref{\Wlocunr}}{(\WPref{\Wrw}{\Wlocunr}{\Whty})^\Wqual}}
  \and \infer []
       {\linsingl{\Wloc}{\Whty} \\
        \Wlin \Wquallessctx \Wqual
       }
       {\valtype{\storetyp}{\funcctx}{\WVref{\Wloclin}}{(\WPref{\Wrw}{\Wloclin}{\Whty})^\Wqual}}
   \and \infer []
       {\linempty \\
         \valtype{\storetyp}{\funcctx}{v}{(\Wpty\subst{\Wptyvar}{\WPrec{\Wptyvar}{\Wqual}{\Wpty^\Wqual}})^\Wqual} \\
         \Wqual \Wquallessctx \Wqual'
       }
       {\valtype{\storetyp}{\funcctx}{\WVfold{v}}{(\WPrec{\Wptyvar}{\Wqual}{\Wpty^\Wqual})^{\Wqual'}}}

  \and \infer []
       { \valtype{\storetyp}{\funcctx}
                 {v}{(\Wpty\subst{\Wlocvar}{\Wloc})^{\Wqual}}  \\
         \Wqual \Wquallessctx \Wqual'
       }
       {\valtype{\storetyp}{\funcctx}{\WVmempack{\Wloc}{v}}{(\WPexists{\Wlocvar}{\Wpty^{\Wqual}})^{\Wqual'}}}
  \and \infer []
       { \linempty \\
         \meminst(i) = M \\
         \modctx.\mathsf{table}(j) = \forall \many{\Wk}. \Waty \\         
       }
       {
         \valtype{\storetyp}{\funcctx}{\WVcoderef{i}{j}{\many{\Wk'}}}
                 {(\Waty\subst{\many{\Wk}}{\many{\Wk'}})^\Wqual}
       }
\end{mathpar}
       }
  \negspace
\caption{Value typing.}\label{fig:valtyp}
\end{figure}

\paragraph{Heap Typing.} The heap typing judgement is written $\heaptype{\storetyp}{\funcctx}{\Whv}{\Whty}$
and asserts that the heap value $\Whv$ has type $\Whty$ in the store
typing $\storetyp$ and function environment $\funcctx$. Complete rules
of heap value typing can be found in the appendix.

\paragraph{Instruction Typing.} The core of \rwasm's type system is typing of instructions. 
The typing judgement, written
$\instrtype{\storetyp}{\modctx}{\funcctx}{\locctx}{\many{e}}{\ftyp}{L'}$
asserts that a list of instructions $\many{e}$ has type
$\ftyp$. $\storetyp$, $\modctx$, and $\funcctx$ are the store typing,
module environment and function environment respectively.
$L$ is the local environment that keeps track of the types and sizes
of the local variables, and $L'$ is the typing of the local variables
after execution of $\many{e}$, which might change the types of local
variables. Many rules follow Wasm with the addition that we add the
necessary premises to ensure linearity (i.e., constraints on
qualifiers and linear memory typing). In Fig. \ref{fig:typing}, we
show several important rules of the system.

\begin{figure}[htb]
  {\scriptsize
    \begin{flushleft}
    \fbox{$\instrtype{\storetyp}{\modctx}{\funcctx}{\locctx}{\many{e}}{\ftyp}{L'}$}
    \end{flushleft}
    \include{instrtyp}
  \negspace
    \caption{Instruction typing.}\label{fig:typing}}
\end{figure}

All block-style instructions follow similar principles, so let's examine
the block instruction $\Wblock{\Wty}{\Wleff}{\many{e}}$.
The block instruction is annotated with the type of the inner list of
instructions $\ftyp$ and the local effects $\Wleff$, which prescribe
the effect that this block of instructions has on the local
environment. The premises of the rule first construct the returned
local environment $L'$, by applying the local effects to the initial
local environment, which is written $\Wleff[L]$ and means that the
type of the $i$th slot of $L$ changes to $(\many{\Wtau})_{(i)}$.
Then the premises assert that block of instructions has the expected
type inside a new function environment that we obtain by pushing
$(\Wty_2, L')$ to the $\mathsf{label}$ component and $\Wunr$ to the
$\mathsf{linear}$ component.
The former tracks the return type and local environment of the label
we are creating. The latter gives us a new qualifier with which we
will track linearity of values on the stack inside this new block and
"locks in" the qualifier corresponding to the linearity of the values
on the stack between the previous enclosing block and this new block.
We choose $\Wunr$ becuase upon entering a block there are no linear
values which might be jumped over. The head element of the $\mathsf{linear}$
environment can be increased by the frame rule (not shown), which as in Wasm, allows
typing rules to ignore values lower on the stack. Since break
instructions will not see ingored values, the only way they can know
whether they will be dropping linear values is by consulting this
environment.

Let's look at some instructions for manipulating locals.
The $\Wgetlocal{i}{q}$ instruction fetches the value of the $i$th
local slot. If the qualifier of the slot, $q$, is linear, then the
contents of the slot must be updated to ensure linearity. We replace
it with the unit value and update the typing of the slot accordingly.
The $\Wsetlocal{i}$ instruction updates the $i$th local slot,
allowing to update its type as well. It ensures that the previous
value was unrestricted, so it can freely be dropped, and that the
upper bound of the size of the new type $\sizeof{\tau}$, fits into the
size of the slot, written $\sizeof{\tau} \Wszlessctx \Wsz$. The size
function takes the $\mathsf{type}$ environment component as parameter
to lookup the upper bound for the size of type variables. The
$\Wszlessctx$ operation takes as parameter the $\mathsf{size}$
component to take into account the size constraints that are in scope.
%
%Lastly, the $\Wteelocal{i}$ instructions allows programs to put a
%value in a local slot while keeping it on the stack. Both the
%qualifier of the previous value which is dropped, and the qualifier of
%the new value which is duplicated, much be unrestricted.

Creating an existential package is done with $\Wmempack{\Wloc}$.
It receives a value from the stack containing
a location $\Wloc$ and creates an existential package that hides this
location.
%% Recall that this instruction reduces to the existential
%% package value $\WVmempack{\Wloc}{v}$ once it receives the value from
%% the stack, whose typing we discussed earlier.
%
The typing rule for $\Wmemunpack{\Waty}{\Wleff}{\Wlocvar}{\many{e}}$
is more complicated. It combines the typing of instructions that
introduce a new block with unpacking
an existential location. The instruction receives from the stack the
arguments of the instruction block $\many{\Wty_1}$ and a packed value
with an existential location type. Then it puts the location variable
in the $\flocation$ environment component to type the instructions in
the block.
The handling of the $\flabel$ and $\flinear$ components is the same as
in the block instruction.

Next we discuss instructions for manipulation heap values. Each family
of heap values has its own malloc instruction. For example,
the $\Wstructmalloc{\manyn{sz}{n}}{q}$ instruction allocates
space for a struct of $n$ values with sizes $\manyn{sz}{n}$. It
receives from the stack $n$ values of types $\manyn{\Wty}{n}$ and it
returns a reference whose location is abstracted with an existential
type and has the corresponding struct type. We require that the size
of the types fit into the requested sizes of the slots. We also put
the restriction that there are no capabilities in the types, for
the reasons described in \S \ref{sec:semantics}.
Structs have get, set, and swap operations shown in the following
rules. $\Wstructget{i}$ gets the $i$th element of a struct, that must
be an unrestricted value, and $\Wstructset{i}$ sets the $i$th element
with the a value that it finds on the stack. The qualifier of the
previous value that is dropped must be unrestricted. If the reference
holding the stack is linear then the type of the new value can be
arbitrary, otherwise it must be the same as the type of the previous
value, as only linear structs support strong updates.
There is also a check the the new value fits in the size of the slot.
The only way to read and write a linear entry from a struct is with a
$\Wstructswap{i}$ operation that combines set and get by
simultaneously getting and setting a struct cell and therefore ensuring
that neither value is dropped or duplicated.

Variants in the heap can be manipulated with the case analysis
instruction. The instruction
$\Wvariantcase{q}{\WHvariant{\manyn{\Wty}{n}}}{\ftyp}{\Wleff}{\manyn{(\many{e})}{n}}$
performs case analysis on a variant with type
$\WHvariant{\manyn{\Wty}{n}}$ and, depending on the result, executes
one block of instructions from the branch list
$\manyn{(\many{e})}{n}$. The rule expects $\many{\Wty_1}$ on the stack
and returns $\many{\Wty_2}$. If the qualifier of the instruction is
linear, then the underlying memory will be freed after the case analysis
and the given (linear) reference will not be returned.
Each of the instruction blocks $(\many{e})_i$ are typed in
the updated (in the usual way) function context and are required to
have type ${\Warrow{\many{\Wty_1}~\Wty_{i}}{\many{\Wty_2}}}$,
where $\Wty_{i}$ is the type of the $i$th variant.
The unrestricted case is similar, with the distinction that the reference
is returned onto the stack after the case analysis and the underlying memory
is left intact.
In addition, the second
element of the $\flinear$ component (corresponding to the values outside
this case block, but inside the nearest surrounding block) is switched to
an arbitrary $q'$ that is stricter than both the qualifier of the variant
reference and the previous qualifier of $\flinear$. This ensures that any
jumps beyond this case block will need to consider whether the reference
we're leaving on the stack is linear, as such a jump woudld drop it.

%% TODO if we need space we can drop this para 
%Lastly, we discuss the administrative instructions for allocating and
%freeing memory.
%%
%The $\Wmalloc{\Wsz}{\Whv}{q}$ instruction allocates a memory slot of
%size $\Wsz$ containing a heap value $\Whv$ in the memory corresponding
%to qualifier $q$.
%%
%It returns an existential package containing a reference to this location.
%It also checks that the heap type does not contian any capabilites, since
%raw capabilities are not allowed on the heap.
%%
%Free takes a linear reference to a heap value and dealocates it. It
%ensures that the qualifier of the reference is linear since
%unrestricted references are managed exclusively by the garbage
%collector.

\paragraph{Configuration and store typing.} At the top level we have
the typing judgements for stores and program configurations, show in Fig. \ref{fig:configtyp}.
The spirit is the same as in Wasm, but we have to split the linear
store typing across the typing of different components of the
configuration. First we introduce an auxiliary judgement
$\conftypefull{\storetyp}{(\many{\Wtau})^?}{i}{\manyn{v}{n};\manyn{\Wsz}{n}}{\many{e}}{\many{\Wtau}}$
that asserts that in the store typing $S$ the configuration
$s;\manyn{v}{n};\manyn{\Wsz}{n};\many{e}$ will result in a stack of type
(or potentially returns) $\many{\Wtau}$. The premises require that
the local values are well-typed with some types $\manyn{\Wty_v}{n}$
and that the size of each (closed) value fits in the size of the
corresponding slot. The instructions have type
$\Wemptyarrow{\many{\Wty}}$ under the empty function context containing
only an optional return type and under the local context
$\manyn{(\Wty_v, sz)}{n}$. The store typing $S$ is the disjoint
union of the store typing used across all typing judgements of the
premises. Furthermore, we require that in the final local environment
there are no linear values, as all of them must be consumed.

The store typing judgement $\vdash s : \storetyp_{\mathit{heap}};
\storetyp_{\mathit{prog}}$ asserts that the store $s$ has typing
$\storetyp_{\mathit{heap}}; \storetyp_{\mathit{prog}}$.  The two store
typings $\storetyp_{\mathit{heap}}$ and $\storetyp_{\mathit{prog}}$
have identical instances and unrestricted memory typings, but disjoint
linear typings.  The linear typing of $\storetyp_{\mathit{prog}}$
contains the surface locations found syntactically in a configration,
i.e., the root pointers, while the linear typing of
$\storetyp_{\mathit{heap}}$ contains the linear locations needed to
type the contents of the memory.
The rule asserts that the domain of the linear memory coincides with
the locations contained in the linear typing of
$\storetyp_{\mathit{heap}}$ and $\storetyp_{\mathit{prog}}$, and the
same for the unrestricted memory.
It also asserts that every pair of location and heap value in the
unrestricted store has the corresponding type prescribed by the
unrestricted memory typing.
For the reasons described previously, it also requires that no capabilities
are present on the heap.
Additionally, $\storetyp_{\mathit{heap}}$ must be the disjoint union
of all the individual store typing components used in typing the heap
components.
Lastly, the rule asserts that the length of the module instances
$|s_{\mathsf{inst}}|$ must coincide with the length of the instance
typings $|S_{\mathsf{inst}}|$ and that each instance in the list has
the corresponding instance typing. We elide the instance typing
judgement from the paper as it is similar to the one of Wasm.

The top level judgement
$\conftype{i}{s}{\manyn{v}{n};\manyn{\Wsz}{n}}{\many{e}}{\many{\Wtau}}$
asserts that the store is well typed in some store typings
$\storetyp_{\mathit{heap}}$ and $\storetyp_{\mathit{proj}}$, and that the
configuration is well typed in the store typing
$\storetyp_{\mathit{proj}}$.

\begin{figure}[htb]
{\scriptsize
\begin{flushleft}
  \fbox{$\conftypefull{\storetyp}{(\many{\Wtau})^?}{i}{\manyn{v}{n};\manyn{\Wsz}{n}}{\many{e}}{\many{\Wtau}}$}
\end{flushleft}
\begin{mathpar}
  \infer []
   { \storetyp = \storetyp_{stack} \uplus \storetyp_1 \uplus \ldots \uplus \storetyp_{n}
     \\ \forall v_i \in \manyn{v}{n}~\Wsz_i \in \manyn{\Wsz}{n},~ \storetyp_i;\funcctx_{empty} \vdash v_i : \Wtau_{v_i} \wedge \| v_i \|_{\epsilon} \Wszless \Wsz_i
     \\ \funcctx = \funcctx_{empty}, \mathsf{return} (\many{\Wtau})^?
     \\ \instrtype{\storetyp_{stack}}{\meminst(i)}{\funcctx}{\manyn{(\Wtau_v, \Wsz)}{n}}{\many{e}}{\epsilon \rightarrow \many{\Wtau}}{L'}
     \\ \forall (\Wpty^\Wqual, \Wsz) \in L',~ \Wqual \Wquallessctx \Wunr}
	{\conftypefull{\storetyp}{(\many{\Wtau})^?}{i}{\manyn{v}{n};\manyn{\Wsz}{n}}{\many{e}}{\many{\Wtau}}}
\end{mathpar}
\begin{flushleft}
  \fbox{$\vdash s : \storetyp_{\mathit{heap}}; \storetyp_{\mathit{prog}}$}
\end{flushleft}
\begin{mathpar}
  \mprset{vskip=2ex}
  \infer []
         {S = \storetyp_{heap} \uplus \storetyp_{prog} \\
           \mathit{dom}~s_{\mathsf{lin}} = \mathit{dom}~\memlin \\
           \mathit{dom}~s_{\mathsf{unr}} = \mathit{dom}~\memunr \\\\
           s_{\mathsf{unr}} = \{ (l_{\Wunr_1}, \Whv_{\Wunr_1}), \dots, (l_{\Wunr_n}, \Whv_{\Wunr_n}) \} \\
           s_{\mathsf{lin}} = \{ (l_{\Wlin_1}, \Whv_{\Wlin_1}), \dots, (l_{\Wlin_m}, \Whv_{\Wlin_m}) \} \\\\ \vspace{4em}
           \forall~i\leq~n,~\heaptype{S_{\mathit{unr}_i}}{\funcctx_{empty}}{\Whv_{\Wunr_i}}{\memunr(l_{\Wunr_i})} \\\\ \vspace{4em}
           \forall~i\leq~m,~\heaptype{S_{\mathit{lin}_i}}{\funcctx_{empty}}{\Whv_{\Wlin_i}}{\memlin(l_{\Wlin_i})} ~\wedge~ l_{\Wlin_i} \in \mathit{codom}~s_{\mathsf{unr}} \Rightarrow \mathit{no\_caps}(\Whv_{\Wlin_i}) \\\\ \vspace{4em}
           \storetyp_{heap} = \biguplus_{i \leq n} S_{\Wunr_i} \uplus \biguplus_{i \leq m} S_{\Wlin_i}
           \\\\ \vspace{4em}
           |s_{\mathsf{inst}}| = |S_{\mathsf{inst}}| \\
           \forall~i\leq\mathit{len},~ S_{\mathsf{inst}} \vdash s_{\mathsf{inst}}(i) : S_{\mathsf{inst}}(i)
         }
         {\vdash s : \storetyp_{\mathit{heap}}; \storetyp_{\mathit{prog}}}
\end{mathpar}
  
\begin{flushleft}
\fbox{$\conftype{i}{s}{\many{v, \Wsz}}{\many{e}}{\many{\Wtau}}$}
\end{flushleft}

\begin{mathpar}
  \infer []
    {\vdash s : \storetyp_{heap}; \storetyp_{prog}
	\\ \conftypefull{\storetyp_{prog}}{\epsilon}{i}{\manyn{v}{n};\manyn{\Wsz}{n}}{\many{e}}{\many{\Wtau}}}
	{\conftype{i}{s}{\manyn{v}{n};\manyn{\Wsz}{n}}{\many{e}}{\many{\Wtau}}}
\end{mathpar}
}
  \negspace
\caption{Configuration typing.}\label{fig:configtyp}
\end{figure}

  \negspace
\subsection{Type safety}
We prove, in Coq, that our language is type safe by proving soundness via progress and preservation.

\paragraph{Progress.}
If a configuration is well-typed
$\conftype{i}{s}{\many{v, \Wsz}}{\many{e}}{\many{\Wtau}}$, then either
$\many{e}$ are all values, or it is a single $\Wtrap$ instruction, or
the configuration can take a step $\Wredi{s}{\many{v}}{\many{\Wsz}}{\many{e}}{s'}{\many{v'}}{\many{e'}}$.

\paragraph{Preservation.}
If a well-formed and well-typed configuration,
$\conftype{i}{s}{\many{v, \Wsz}}{\many{e}}{\many{\Wtau}}$, takes a step
$\Wredi{s}{\many{v}}{\many{\Wsz}}{\many{e}}{s'}{\many{v'}}{\many{e'}}$
then the resulting configuration is well-typed with the same type
$\conftype{i}{s'}{\many{v', \Wsz}}{\many{e'}}{\many{\Wtau}}$

\paragraph{Coq development.} We have formalized the language, its static and dynamic semantics,
and the proof of type safety via progress and preservation in Coq.
The effort is substantial, consisting of ~14k lines of specifications
(definitions and theorem statements) and ~52k lines of proofs, all
directly related to the type system.
We have submitted the proof development as supplemental material. 
At the time of the submission we have 4 remaining admitted lemmas,
all related to substitution (among many others that we have fully proved).
We give a more detailed description in the submitted artifact.

%% TODO I cannot find these in Coq for configurations.
  \negspace
\subsection{Example}

We conclude this section by considering some examples to see how \rwasm's fine-grained
memory access might be useful for real programs. Imagine a library for some performance-critical
operation, such as a graphics library. We  want to implement this in a manually managed
source language. Instances of this graphics data structure might take some mutable configuration
state that can change over the course of a program, such as quality settings or dimensions.
Next, we might want to write the higher-level logic of our program in a GC'd language which simply
makes use of this library for graphics. Such a program requires the GC'd code to reference linear
values, which in turn reference some shared mutable state.

For our example, we'll keep this structure, but simplify our library down to a small mutable counter.
The shared runtime state will configure how much counters should increment by. The GC'd portion of
our program will use this linear library, but hide it behind an interface which allows it to use
the library without reasoning about linearity.

%We've shown that \rwasm's types allow for the linear and unrestricted memories to reference
%each other in a fine grained manner, but let's consider some examples to see how this might be
%useful to real programs. First, imagine some library for some performance-critical operation.
%We would want to implement this in a manually managed source language. For our small
%example, we will use a simple stateful counter, but one could imagine having a large linear
%data structure which underlies a graphics library. If we have multiple of these data
%structures and they are configurable in some way, then we might want shared mutable state to
%control them all. In the small example, this will be how much our counter should increment by,
%but in a graphics library one could imagine settings for quality or dimensions, which may need
%to change as the program runs. Lastly, we might want to write the higher level logic of our
%program in a garbage collected language, using this performant library to get some computational
%result. In the small example, we'll use existential types to hide the underlying type of the
%linear data structure and expose only a function which increments the counter and returns the
%result.

\begin{figure}[htb]
  \begin{minipage}[b]{0.41\textwidth}
    \flushleft
    \includegraphics[width=\textwidth]{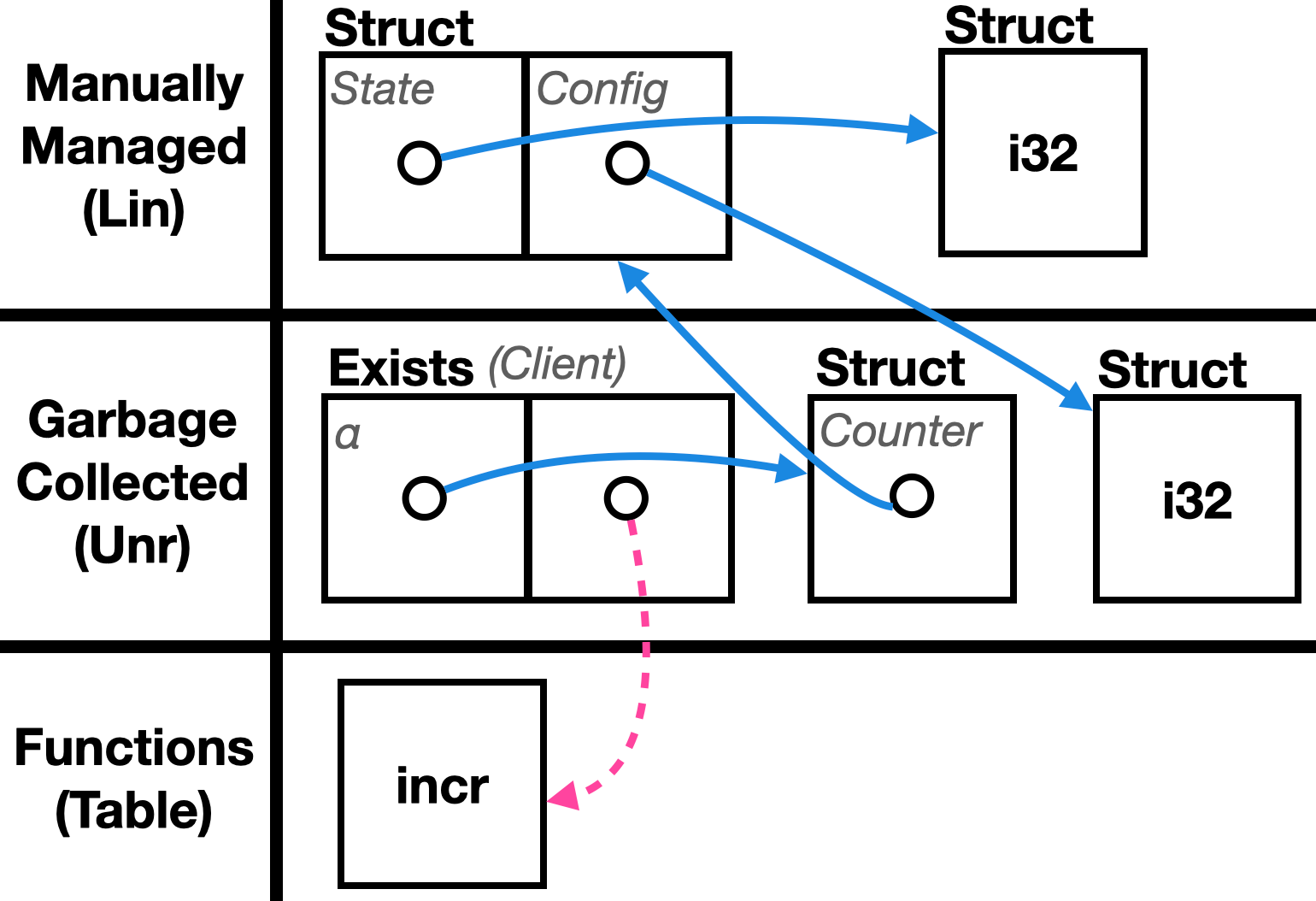}
  \end{minipage}
  \begin{minipage}[b]{0.58\textwidth}
    \begin{scriptsize} \hspace{2mm}
    \begin{align*}
\textsf{Client} & := 
\WHexists{\Wunr}{\Wptyvar}{64}{\WPgroup{\Wptyvar^{\Wunr}
    (\WPcoderef{\forall~\epsilon.~\Wptyvar^{\Wunr} \rightarrow \Wkey{i32}^{\Wunr}})^{\Wunr}}^{\Wunr}} \\
\textsf{$\alpha$} & := 
  (\WPexists{\ell}{(\WPref{\ell}{\Wrwrw}{\WHstruct{Counter,~64}})^{\Wunr}})^{\Wunr} \\
\textsf{Counter} & := 
(\WPexists{\ell}{(\WPref{\ell}{\Wrwrw}{\WHstruct{(State,~64) 
        (Config,~64)}})^{\Wlin}})^{\Wlin} \\
\textsf{State} & :=
  (\WPexists{\ell}{(\WPref{\ell}{\Wrwrw}{\WHstruct{\Wkey{i32}^{\Wunr},~32}})^{\Wlin}})^{\Wlin} \\
\textsf{Config} & :=
(\WPexists{\ell}{(\WPref{\ell}{\Wrwr}{\WHstruct{\Wkey{i32}^{\Wunr},~32}})^{\Wunr}})^{\Wunr} \\
\\
\end{align*}
  \end{scriptsize}
  \end{minipage}
  \negspace
 \caption{Memory layout. (ref = solid blue line) (coderef = dashed pink line) and type definitions.}\label{fig:ex2}
\end{figure}

%% \begin{subfigure}[b]{0.4\textwidth}
%% \flushleft
%% \includegraphics[width=\textwidth]{figure_2_picture.png}
%% \subcaption{Memory layout. (ref = solid blue line) (coderef = dashed pink line)}\label{subfig:ex2}
%% \end{subfigure}
%% %% \hspace{0.1em}
%% \begin{subfigure}[b]{0.59\textwidth} 
%%   \begin{scriptsize} \hspace{2mm}
%%     \begin{align*}
%% \textsf{Client} & := 
%% \WHexists{\Wunr}{\Wptyvar}{64}{\WPgroup{\Wptyvar^{\Wunr}
%%     (\WPcoderef{\forall~\epsilon.~\Wptyvar^{\Wunr} \rightarrow \Wkey{i32}^{\Wunr}})^{\Wunr}}^{\Wunr}} \\
%% \textsf{$\alpha$} & := 
%%   (\WPexists{\ell}{(\WPref{\ell}{\Wrwrw}{\WHstruct{Counter,~64}})^{\Wunr}})^{\Wunr} \\
%% \textsf{Counter} & := 
%% (\WPexists{\ell}{(\WPref{\ell}{\Wrwrw}{\WHstruct{(State,~64) 
%%         (Config,~64)}})^{\Wlin}})^{\Wlin} \\
%% \textsf{State} & :=
%%   (\WPexists{\ell}{(\WPref{\ell}{\Wrwrw}{\WHstruct{\Wkey{i32}^{\Wunr},~32}})^{\Wlin}})^{\Wlin} \\
%% \textsf{Config} & :=
%%   (\WPexists{\ell}{(\WPref{\ell}{\Wrwr}{\WHstruct{\Wkey{i32}^{\Wunr},~32}})^{\Wunr}})^{\Wunr}
%% \end{align*}
%%   \end{scriptsize}
%%  \subcaption{Type definitions. \\ $~$ \\ $~$ }\label{subfig:ex2}
%% \end{subfigure}
%%  %% \caption{Counter example.}\label{fig:ex2}
%% \end{figure}

Fig. \ref{fig:ex2} presents a memory layout for such an example.
\textsf{Client} is the value with which our GC-ed portion will be interacting.
It contains a pair of an abstract value $\Wptyvar$ and a coderef which, given that value,
increments the counter and returns the new count.
The hidden type $\Wptyvar$ contains a struct referencing \textsf{Counter}, the main data
structure provided by our linear library.
\textsf{Counter} contains a pair of references, one which grants write access ($\Wrwrw$)
to its internal mutable state (\textsf{State}), and one which grants read access ($\Wrwr$)
to the shared configuration (\textsf{Config}).
Once this heap is laid out, the GC'd portion of the program can configure and use
the counter without any need to reason about linearity at all.
A program which creates this heap layout is included in our supplemental material.

%Figure \ref{fig:ex2} presents a memory layout of our counter use
%case. Note that the portion contained in the gray rectangle does not
%exist concurrently with the rest of the image but represents a
%different point in time.
%
%\textsf{Config} is a reference to the global configuration shared by
%all counters. \textsf{State} is a read-only reference to the internal
%state of the counter. These are put together into a data structure
%called a \textsf{Counter}, the primary type provided by the counter
%libarry. Note that \textsf{State} is owned by the garbage collector
%(as it its capability currently lives in the unrestricted memory), so
%it may not contain raw capabilities. During certain points in program
%execution this type will not be owned by the garbage collecor, for
%instance when the counter library's functions are operating on
%it. During these times, it is free to place capabilities on the heap,
%as in \textsf{Lin\_state}.  Thus, the counter library only needs to
%worry about this restriction at the boundary with the GC-ed program
%because it might try to place the data structure in the custody of the
%gc.  \textsf{Client} is the value with which our GC-ed program will be
%interacting. It contains an abstract data structure and a coderef to
%the increment function previously described.

%%% Local Variables:
%%% mode: latex
%%% TeX-master: "main"
%%% End:

%% file: instrtyp.tex
\begin{mathpar}
  %% %
  %% % Value
  %% %      
  %% \infer []
  %%        {\valtype{\storetyp}{\funcctx}{v}{\tau}}
  %%        {\instrtypectx{v}{\Wemptyarrow{\tau}}{\locctx}}
  %
  % Block
  %      
  \and \infer []
       { \locctx' = \Wleff[\locctx] \\
          \funcctx' = \funcctx, \mathsf{label}~(\Wty_2, \locctx') :: \flabel, \mathsf{linear}~\Wunr :: \flinear \\
         \instrtype{\storetyp}{\modctx}{\funcctx'}{\locctx}{\many{e}}{\Warrow{\many{\Wty_1}}{\many{\Wty_2}}}{\locctx'} \\
        }
        {\instrtypectx{\Wblock{\Warrow{\many{\Wty_1}}{\many{\Wty_2}}}{\Wleff}{\many{e}}}
         {\Warrow{\many{\Wty_1}}{\many{\Wty_2}}}
         {\locctx'}}
  %
  % GetLocal
  %      
  \and \infer []
        { \linempty \\
          \locctx_{(i)} = (\Wpty^{\Wqual}, \Wsz) \\
          (\Wqual \Wquallessctx \Wunr \wedge \Wtau' = \Wpty^{\Wqual}) \vee
          (\neg \Wqual \Wquallessctx \Wunr \wedge \Wtau' = \WPunit^{\Wqual})
          \\
          %% \qualvalid{\funcctx}{\Wqual}
        }
        {\instrtypectx{\Wgetlocal{i}{\Wqual}}
          {\Wemptyarrow{\Wpty^{\Wqual}}}{\locctx[i \mapsto(\Wtau', \Wsz)]}}
  %
  % SetLocal
  %      
  \and \infer []
        { \linempty \\
          \locctx_{(i)} = (\Wpty^{\Wqual}, \Wsz) \\
          \Wqual \Wquallessctx \Wunr \\
          \sizeof{\tau} \Wszlessctx \Wsz \\
        }
        {\instrtypectx{\Wsetlocal{i}}
          {\Wemptyres{\Wtau}}{\locctx[i \mapsto(\Wtau, \Wsz)]}}
  %
  % TeeLocal
  %      
%  \and \infer []
%        { \linempty \\
%          \locctx_{(i)} = (\Wpty^{\Wqual}, \Wsz) \\
%          \Wqual \Wquallessctx \Wunr \\
%          \Wtau = \Wpty'^{\Wqual'} \\
%          \Wqual' \Wquallessctx \Wunr \\
%          \sizeof{\tau} \Wszlessctx \Wsz \\
%        }
%        {\instrtypectx{\Wteelocal{i}}
%          {\Warrow{\Wtau}{\Wtau}}{\locctx[i \mapsto(\Wtau, \Wsz)]}}
  % 
  % mempack
  %
  \and \infer []
       {\linempty \\ }
       {\instrtypectx{\Wmempack{\Wloc}}
         {\Warrow{(\Wpty\subst{\Wlocvar}{\Wloc})^\Wqual}
           {(\WPexists{\Wlocvar}{\Wpty^\Wqual})^\Wqual}}
           {\locctx}}
  %
  % memunpack       
  %
  \and \infer []
       {
         \locctx' = \Wleff[\locctx] \\
         \funcctx' = \funcctx, \mathsf{label}~(\many{\Wty_2}, \locctx') :: \flabel,
                     \mathsf{linear}~\Wunr :: \flinear,
		     \mathsf{location}~\Wlocvar :: \flocation
         \\
         \instrtype{\storetyp}{\modctx}{\funcctx'}{\locctx}{\many{e}}
                   {\Warrow{\many{\Wty_1}~\Wty_\rho}{\many{\Wty_2}}}
                   {\locctx'} \\
       }
       {\instrtypectx
	 {\Wmemunpack{\Warrow{\many{\Wty_1}}{\many{\Wty_2}}}{\Wleff}{\Wlocvar}{\many{e}}}
         {\Warrow{\many{\Wty_1}~(\WPexists{\Wlocvar}{\Wty_\rho})^{\Wqual_\rho}}
           {\many{\Wty_2}}}{\locctx'}}
  %       
  % Struct malloc 
  %
  \and \infer []
       { \linempty \\
         \nocaps~\manyn{\Wtau}{n}\\ %% TODO add heapable
         \forall~(\tau_i, \Wsz_i) \in \manyn{(\Wty, \Wsz)}{n},~\sizeof{\tau_i} \Wszlessctx \Wsz_i \\
        }
       { \instrtypectx{\Wstructmalloc{\manyn{sz}{n}}{\Wqual}}
         {\Warrow{\manyn{\Wtau}{n}}
           {\WPexists{\Wlocvar}{(\WPref{\Wrwrw}{\Wlocvar}{\WHstruct{\manyn{(\Wty, \Wsz)}{n}}})^{\Wqual}}}}
           {\locctx}
       }       
  %
  % struct get 
  %
  \and \infer []
       { \linempty \\
         \many{\Wty}_{(i)} = \Wpty_i^{\Wqual_i} \\
         \Wqual_i \Wquallessctx \Wunr \\
        }
       {\instrtypectx{\Wstructget{i}}
         {\Warrow{(\WPref{\Wrw}{\Wloc}{\WHstruct{\many{(\Wty, \Wsz)}}})^\Wqual}
           {(\WPref{\Wrw}{\Wloc}{\WHstruct{\many{(\Wty, \Wsz)}}})^\Wqual; \Wpty_i^{\Wqual_i}}}
         {\locctx'}}
 %       
  % Struct set 
  %
  \and \infer []
        { \linempty \\
         \many{\Wty}_{(i)} = \Wpty_i^{\Wqual_i} \\
         \Wqual_i \Wquallessctx \Wunr \\         
         \sizeof{\tau_i'} \Wszlessctx \many{\Wsz}_{(i)} \\
         \nocaps~{\Wtau_i'}\\
           \Wlin \Wquallessctx \Wqual \vee \Wtau_i' = \Wpty_i^{\Wqual_i} \\
        }
        {\instrtypectx{\Wstructset{i}}
         {\Warrow{(\WPref{\Wrw}{\Wloc}{\WHstruct{\many{(\Wty, \Wsz)}}})^\Wqual; \Wty_i'}
           {(\WPref{\Wrw}{\Wloc}{\WHstruct{\many{(\Wty[i \mapsto \Wty_i'], \Wsz)}}})^\Wqual}}
         {\locctx'}}
   %
   % struct swap 
   %
   \and \infer []
        { \linempty \\
         (\many{\Wty})_{(i)} = \Wpty_i^{\Wqual_i} \\
         \sizeof{\tau_i'} \Wszlessctx \many{\Wsz}_{(i)} \\
         \nocaps~{\Wtau_i'}\\
         \Wlin \Wquallessctx \Wqual \vee \Wtau_i' = \Wpty_i^{\Wqual_i} \\
        }
        {\instrtypectx{\Wstructswap{i}}
         {\Warrow{(\WPref{\Wrw}{\Wloc}{\WHstruct{\many{(\Wty, \Wsz)}}})^\Wqual; \Wty_i'}
           {(\WPref{\Wrw}{\Wloc}{\WHstruct{\many{(\Wty[i \mapsto \Wty_i'], \Wsz)}}})^\Wqual}}
         {\locctx'}}
        \\
        %
        % variant case linear
        %
        \and \infer []
             { \linempty \\
                \locctx' = \Wleff[\locctx] \\
                \funcctx' = \funcctx, \mathsf{label}~(\many{\Wty_2}, \locctx') :: \flabel,
                \mathsf{linear}~\Wunr :: \flinear \\         
                \forall~i\leq~n,~\instrtype{\storetyp}{\modctx}{\funcctx'}{\locctx}{(\many{e})_{(i)}}{\Warrow{\many{\Wty_1}~(\many{\Wty})_{(i)}}{\many{\Wty_2}}}{\locctx'} \\
                \Wlin \Wquallessctx \Wqual \\
                \Wlin \Wquallessctx \Wqual_v \\
                \Whty = \WHvariant{\manyn{\Wty}{n}}
             }
             { \instrtypectx{\Wvariantcase{q}{\Whty}{\ftyp}{\Wleff}{\manyn{(\many{e})}{n}}}
               {\Warrow{(\WPref{\Wrw}{\Wloc}{\Whty})^{q_v}~\many{\Wty_1}}
                 {\many{\Wty_2}}}
               {\locctx'}
             }
    %
        %
        % variant case unrestricted
        %
        \and \infer []
             { \linempty \\
                \Wqual_v \Wquallessctx \Wqual' \\
                get\_hd ~\flinear \Wquallessctx \Wqual' \\
                \locctx' = \Wleff[\locctx] \\
                \funcctx' = \funcctx, \mathsf{label}~(\many{\Wty_2}, \locctx') :: \flabel,
                \mathsf{linear}~\Wunr :: set\_hd~\Wqual'~\flinear \\
                \forall~i\leq~n,~\instrtype{\storetyp}{\modctx}{\funcctx'}{\locctx}{(\many{e})_{(i)}}{\Warrow{\many{\Wty_1}~(\manyn{\Wty}{n})_{(i)}}{\many{\Wty_2}}}{\locctx'} \\
                \forall ~ {p_i}^{q_i} \in \manyn{\Wty}{n}, ~ \Wqual_i \Wquallessctx \Wunr \\
                \Wqual \Wquallessctx \Wunr \\
                \Whty = \WHvariant{\manyn{\Wty}{n}}
             }
             {\instrtypectx{\Wvariantcase{q}{\Whty}{\ftyp}{\Wleff}{\manyn{(\many{e})}{n}}}
               {\Warrow{(\WPref{\Wrw}{\Wloc}{\Whty})^{q_v}~\many{\Wty_1}}
                 {(\WPref{\Wrw}{\Wloc}{\Whty})^{q_v}~\many{\Wty_2}}}
               {\locctx'}
             }
    % malloc 
    %
        \and \infer []
             {
               \heaptype{\storetyp}{\funcctx}{\Whv}{\Whty} \\
               \qualvalid{F}{q} \\               
               \nocaps~\Whty\\ %% TODO add heapable
             }
             {\instrtypectx{\Wmalloc{\Wsz}{\Whv}{q}}{\Wemptyarrow{(\WPexists{\Wlocvar}{(\WPref{\Wrwrw}{\Wlocvar}{\Whty})^q})^q}}{\locctx}}
             \\
    %
    % free 
    %
        \and \infer []
             {
               \linempty \\
               \Wlin \Wquallessctx q \\               
               \typevalid{F}{(\WPref{\Wrwrw}{\Wloc}{\Whty})^q} \\
               %%      HeapTypeUnrestricted F ht -> TODO what is this
             }
             {\instrtypectx{\Wfree}{\Wemptyres{(\WPref{\Wrwrw}{\Wloc}{\Whty})^q}}{\locctx}}
    %
    % frame 
    %
        \and \infer []
             {
               q_{hd} = \gethd~\flinear \\
               q_f \Wquallessctx q_{hd} \\               
               \forall~{p_i}^{q_i} \in \many{\Wtau}, ~  q_i \Wquallessctx q_f \\
               \funcctx' = \funcctx, \mathsf{linear}~set\_hd~\Wqual_f~\flinear \\
               \forall~\Wtau \in \many{\Wtau},~\typevalid{F}{\Wtau} \\
               \instrtypectx{\many{e}}
                            {\Warrow{\many{\Wtau_1}}{\many{\Wtau_2}}}
                            {\locctx'}
             }
             { \instrtypectx{\many{e}}
                 {\Warrow{\many{\Wtau}~\many{\Wtau_1}}{\many{\Wtau}~\many{\Wtau_2}}}
                 {\locctx'}}
  %
  % exists pack TODO
  %
  %
  % exists unpack TODO
  %
\end{mathpar}

%% file: compilers.tex
  \negspace
\section{Compiling to \rwasm}
\label{sec:compilers}

In order to demonstrate that \rwasm provides a reasonable target for a
variety of high-level languages, we implement type-preserving compilers
from a garbage collected language (\ml) and a manually managed language
(\lthree). In order to demonstrate \rwasm's ability to serve as a
platform for interop between languages, we extend \ml and \lthree with
the necessary constructs to reference each other's types in a limited
fashion. For these extensions, we follow a linking types \cite{} approach,
which aims to allow users of a source language to link with other types
inexpressible in their own language, without losing native reasoning principles.
Additionally, programs which do not mention any extensions are unaffected
by the extensions' presence in the language.

\paragraph{\ml.} Our base \ml supports the standard types: units, ints, references,
variants, products, recursive types, and functions with parametric polymorphism.
Since \rwasm has similar types, the choice of
representation is quite straightforward.
We also extend \ml with standard constructs for writing multi-module code, such as
function imports and exports and the ability to define global state which exported
functions can close over.
Such extensions provide a good basis for exploring multi-module code in \rwasm and
eventually interoperability with other languages.

As described in the discussion of Fig. \ref{fig:full_example_1}, we extend \ml with
the ability to direct the compiler to compile particular types as linear and provide
a construct which allows the creation of references containing linear types.
The \ml compiler explicitly does not check whether types annotated as linear are used
linearly, as we can rely on \rwasm to demonstrate safety. The goal of interoperability
is to allow programmers to use the right language for each task, without burdening them
by turning the type system of their source into something as complex as \rwasm. Programmers
are still writing \ml, only using these linking types at the boundary between languages.

\paragraph{\lthree.} We compile the core \lthree language with a minor adjustment.
We require that \lthree capabilities explicitly track the size of the memory they reference.
Pragmatically, sizes will need to be reasoned about at the \rwasm level, but also
philosophically, a source language which allows precise reuse of memory \emph{should}
require a programmer to think about sizes.
While \lthree's ability to perform strong updates is impressive, it must be accompanied
by reasoning about the size of the location being updated to be useful.

As described for the example in Fig. \ref{fig:full_example_1}, we extend \lthree with an \ml-like reference \textsf{Ref}
type in order to allow memory interop between the two languages, as well as constructs (\textsf{join}, \textsf{split}) to
convert between capability-pointer pairs and references.

Readers familiar with \lthree might note that its source programs often put capabilities on
the heap, an operation which is disallowed in the verion of \rwasm presented in this paper.
While pointers and capabilites can be separated, \lthree programs can always place
capabilities on the heap with some pointer in the form of a reference without losing the
performance gains that come from moving \emph{raw} capabilities around on the stack.
%
%This is however, a bit unsatisfying, as we need to use our "linking" types in the core of
%our language.
%%
%As such, 
Nonetheless, we've been working on relaxing this restriction in \rwasm such that capabilities
are only disallowed in the parts of the heap owned by the garbage collector.
We're in the process of updating our formalism, but our prototype compilers and \rwasm 
typechecker already support this more relaxed restriction, allowing programmers to write
\lthree in their native style.

\paragraph{Compilation.}
Compilation to untyped or poorly typed targets requires compiler writers to
expend significant effort on low-level details like data layout, calling
conventions, and bit manipulation. In contrast, compilation to \rwasm is
largely an exercise in satisfying \rwasm's type system. While this is not
entirely straightforward, we have found it to be tractable.

The \ml compiler has three phases of note: typed closure conversion,
annotation, and code generation.
Typed closure conversion is standard in compilers for functional languages.
What is notable is that (after the remaining passes) \rwasm is going to check
the correctness of this pass, examining the size and qualifier annoations
that we use when generating existential packages to hide the environments of
closures.
Since all type variables in \rwasm have size and qualifier bounds and \ml has
no such notion at the source, we introduce an annotation phase to change the
types of functions appropriately, annotating definitions and call sites with
the appropriate information.
Code generation is overall quite similar to code generation to any stack based
language.
However there are some type-based complications.
For instance, instead of relying on a compiler writer's carefulness when
shuffling values around, \rwasm requires explanation of one's use of locals
in the form of a local effect annotation on every block of
code, allowing it to check that duplication only occurs when allowed.
% 
%Something we would like to explore in the future is how we might compile more
%efficiently by, for example, using local space more optimally.
%
%In order to take advantage of \rwasm's ability to reason about reuse, we would
%need to do additional inference on \ml programs to generate additional constraints
%on our \rwasm types.

\lthree is a much lower level language, without the ability to existentially or
universally quantify types. Thus, it is much easier to compile to \rwasm, and we
can do so in one code generation phase.
To keep things simple and since we've already demonstrated how to do so in \ml, we
don't implement closure conversion in \lthree.

%% file: lowering.tex
  \negspace
\section{Compiling \rwasm to Wasm}
\label{sec:lowering}

Compilation from \rwasm to WebAssembly is type directed.
It requires some type information that is implicit in \rwasm instructions which is provided by the type checker.
%
%% \zoe{I think this sentence gives little information}
%% Instructions that need these explicit type annotations include operations on local and global variables and structs and Arrays, to help the compiler reason about the shape of the Wasm stack before and after an operation. 
% 
The \rwasm to Wasm compiler takes in the type annotated \rwasm code produced by the \rwasm type checker and compiles to WebAssembly 1.0 with the multi-value extension, where, functions and instructions can return more than one value. 

\paragraph{Lowering \rwasm's Type System.}
Every \rwasm type will be translated to a series of base Wasm numeric types.
% PreTypes 
Types with no runtime information, such as $\Wkey{unit}$, $\Wkey{cap}$ and $\WPown$, are erased.
Numeric types are translated to the corresponding WebAssembly  type.
$\Wkey{ref}$  and  $\Wkey{ptr}$ are lowered to a single $\Wkey{i32}$ pointer. 
Type variables $\Wptyvar$ are annotated with a size bound that indicates the maximum size of the type that the $\Wptyvar$ can be instantiated with. 
If this size is concrete or an upper bound for it can be inferred using the constraints, we compile the type into a series of Wasm's numeric types.
%% $\Wkey{32}$ and $\Wkey{64}$ bits to hold values of that type.
%% For example, if an $\Wptyvar$ is bound with a concrete size 160, we lower it to a series of two $\Wkey{i64}$s and an $\Wkey{i32}$.
If the upper bound is unknown, the $\Wptyvar$ is boxed on the heap and it is translated to an $\Wkey{i32}$ pointer.
%% If the upper bound is known, we lower the $\Wptyvar$ to a series of WebAssembly numeric types, as we did when the $\Wptyvar$ was bound with a concrete size.
%
To translate recursive types, we just compile the inner type, since \rwasm gives us an invariant that the recursive type appears only inside a level of indirection.
For existential types, we similarly lower the inner type down to Wasm numeric types. 

\paragraph{Operations on local and global variables.}
Local variables in \rwasm can have arbitrary sizes and can have multiple types over their lifetime.
In Wasm, however, local variables can only be one of four numeric types.
Therefore, we lower a \rwasm local to a series of Wasm locals.
For example, if a \rwasm local has size 160 it will be stored across three Wasm locals of types $\Wkey{i64}$, $\Wkey{i64}$ and $\Wkey{i32}$.
This sequence of locals might be used to store any type of size up to 160, for example $\WPgroup{\Wkey{i32}~\Wkey{i64}~\Wkey{i64}}$.
Therefore, the first local will store the first $\Wkey{i32}$ component of the tuple and the first half of the second $\Wkey{i64}$ component.
The compilation of $\Wkey{local.set}$ and $\Wkey{local.get}$ needs to perform the correct accesses to fetch the entire value onto the stack, and is informed by the \rwasm type.
Operations on global variables in \rwasm is similar to locals. 
% later mention: We choose a naive representation that can lead to a lot of shuffling around of bits. But with some static analysis, we can do better packing of types. 

\paragraph{Memory model and heap types.}
In Wasm we use only one flat memory to represent both \rwasm's memories.
We use a simple free list allocator to allocate and free pointers in
Wasm memory.
%
% Structs 
Structs and arrays are encoded in the Wasm memory as a consecutive bytes. Similar to local variables, the representation of a field or an array slot might need to use more than one consecutive memory slot.
%
%% laid out in memory as consecutive values
%% When allocating space for a struct during a $\Wkey{struct.malloc}$, we allocate the sum of the size of the types of each field in the struct and initialize the fields. Since our compilation is type directed, the $\Wkey{struct.get}$, $\Wkey{struct.set}$ and $\Wkey{struct.swap}$ operations read and write to the heap as directed by the type of the struct field. 
%% % Arrays 
%% Compilation of \rwasm arrays is very similar to structs, except that the number of arrays slots and the index into an array is only known at runtime.
%% Hence, an $\Wkey{array.malloc}$ computes the total size of the array at runtime and all operations on arrays, like, $\Wkey{array.get}$ and $\Wkey{array.set}$, read and write according to the type of an array cell.   
%% \paragraph{Variants and existential types}
Variants are represented in memory as a sequence of bytes containing the numeric tag followed by the corresponding type.
%
%% When lowering the $\Wkey{variant.malloc}$ instruction, we know the types associated with each tag of the variant and the tag of the current variant. Hence, we load into memory the tag and associated data.
%
$\Wkey{variant.case}$ instructions are compiled as a switch case for every case in the variant. Switch cases are represented in Wasm using nested blocks, with blocks for every case, and a $\Wkey{br\_table}$ instruction in the innermost block that jumps to the case being executed. At the start of every case, we provide instructions to read data from the heap according to the type of that case.
$\Wkey{exists.pack}$ stores a single \rwasm value on the heap, and $\Wkey{exists.unpack}$ reads it, with the help of an annotation of the type.
%% represented by the existenial on the heap,
%% takes values on the stack and places them on the heap, according to the shape of the witness type of the existential. We annotate $\Wkey{exists.unpack}$ with the witness type and use it to read from the heap before translating the body of the $\Wkey{exists.unpack}$.

\paragraph{Function calls.}
Functions in \rwasm can be polymorphic on types with unknown size bounds that are represented in Wasm as $\Wkey{i32}$ pointers.
%% For instance, in the type $\Wptyvar$ with a unknown size bound. 
%% Since the $\Wptyvar$ has no size bound, it is boxed on the heap and lowered to a $\Wkey{i32}$ pointer.
%
%% When such a function is called, these $\Wptyvar$'s will be instantiated with some concrete type. 
Say that a caller needs to pass an argument of type
$\WPgroup{\Wkey{i64}~\Wkey{i32}}$ to a function that expects a boxed
representation of the same argument as it is polymorphic on its type.
The caller will put an $\Wkey{i64}$ and $\Wkey{i32}$ on the stack.
Then, we need to perform a stack coercion, replacing the $\Wkey{i64}$
and $\Wkey{i32}$ with a pointer to the same data on the heap.  Stack
coersions like this will always be required when functions expect or
return values of boxed $\Wptyvar$ types.

For indirect calls, $\Wkey{coderef}$ instructions compile to an
$\Wkey{i32}$ index into the function table.
%
%% $\Wkey{inst}$ instructions are erased since they perform type level
%% operations.
%
To coerce the stack to the shape that the callee expects, we make a
case for each possible shape in the table that could correspond to this call,
and at runtime we jump to the correct case depending on the value of the
index to the table.
%% For indirect calls, the stack of the caller needs to be coerced to the
%% stack that the callee expects, but we do not know statically which
%% function will be called. However, we do know all functions that could
%% be indirectly called since the indirect call has a static type, as do
%% all functions in the table. Hence, we use Memory Modelthe index value into the
%% table to specialize the stack for an indirect call.

\paragraph{Remaining Instructions.}
Instructions that have identical counterparts in Wasm are left
unaltered. $\Wrecfold$, $\Wrecunfold$, $\Wmempack{\Wloc}$,
$\Wseqgroup$, $\Wsequngroup$, $\Wcapsplit$, $\Wcapjoin$,
$\Wrefdemote$, $\Wrefsplit$, $\Wrefjoin$, $\Wqualify$, $\Wkey{inst}$ are all erased
since they are type level operations.
%% $\Wkey{mem.unpack}$ is effectively erased and is compiled down to a
%% block instruction.

%\paragraph{Future Work}
%\michelle{Haven't gone over this yet - will do tomorrow.}
%% Drawbacks/where can we improve? 
%There are several optimizations that can be done to improve
%performance of the compiler. However, this has not been our focus and
%we hope to explore it in future work. For example, we can efficiently
%track qualifiers to assist with garbage collection\zoe{elaborate}, or optimize the
%local type layout that we choose when packing values into WebAssembly
%local variables. We could also do dead code elimination to erase empty
%functions that just shuffle locals or inline functions that are mostly doing type system operations.\zoe{Q: are there optimization passes for Wasm that could be possibly reused.?}

%
%%% Local Variables:
%%% mode: latex
%%% TeX-master: "main"
%%% End:

%% file: related.tex
  \negspace
\section{Related Work}
\label{sec:related}

In \S\ref{sec:intro}, we've already discussed the three most closely related
piece of work: \lthree, the Component Model, and Patterson et al.'s semantic
framework for sound interoperability. In a general sense, \rwasm is also
influenced by work on substructural types, using $\Wunr$ and $\Wlin$ qualifiers
to annotate pretypes as in~\cite{ahmed05:substruct}, and by work on
type-preserving compilation and typed low-level
languages~\cite{minamide96,morrisett99:toplas,morrisett02,hawblitzel07:gctal}.

Wasm does not currently support garbage collection natively, but a 
proposal to do so is currently working its way through the standardization
process~\cite{wasmgc}. \rwasm's heap types are intentionally designed
to be compatible with this proposal, but we believe it lacks one crucial
feature: finalizers. Many languages use finalizers, and for \rwasm they
are essential to allowing the garbage collector to own, and thus sometimes
free, linear memory. At present, with Wasm's current GC proposal, \rwasm's
runtime needs to implement its own garbage collector.

MSWasm~\cite{mswasm} (Memory-Safe Wasm) is an extension of Wasm designed to
enforce memory-safe execution of unsafe code, e.g., code compiled from C
or C++. MSWasm extends Wasm with language constructs for CHERI-like fine-grained
dynamically checked memory capabilities so code compiled from C will be checked
for memory safety at runtime. In contrast, \rwasm has static rather than dynamic
capabilities, which have the benefit of zero runtime overhead. But \rwasm is
meant to be a target for type-safe source 
languages; when compiling an unsafe language like C it would be nearly
impossible to produce type-annotated \rwasm code that is well typed since the 
type information and safety guarantees don't exist in the source.

Iris-Wasm~\cite{iriswasm} is a mechanized higher-order separation logic for
modularly verifying Wasm 1.0 programs. The authors have used Iris-Wasm to build
a logical relation for Wasm and prove type safety. By contrast, we have a
mechanized type safety proof for a language with a far richer type system, but
\rwasm is a typed language not a verification logic. 

% \item Type safe FFIs~\cite{furr08}

%%% Local Variables:
%%% mode: latex
%%% TeX-master: "main"
%%% End:

%% file: discussion.tex
\section{Discussion and Future Work}
\label{sec:discussion}

We want \rwasm to serve as a platform for safe interoperability between a wide range of typed languages, which may require extending its type system. 
Our first priority in future work is type-preserving compilation from safe Rust  to \rwasm. There are two interesting challenges: how to encode lifetime constraints and how to encode immutable borrows in \rwasm. The latter may require fractioal capabilities so we can create many borrows but when all of these borrows end, we can produce a linear capability to return to the owner.  A further challenge is to compile Rust with concurrency to a concurrent extension of \rwasm, in turn compiling that to the recent Wasm threads proposal. 

Next, we will tackle compiling languages with advanced control effects, e.g., algebraic effect handlers. This will require extending \rwasm with linearly typed continuations and compiling \rwasm to Wasm with the recent typed continuations proposal~\cite{wasmfx,wasmfx:oopsla}. The latter dynamically ensures that continuations are used linearly instead of using linear types, which are expensive to implement as they preclude in-place updates. In \rwasm, we can statically ensure linear usage of continuations and then have the compiler to Wasm perform the optimization to use in-place updates. We can verify correctness of such optimizations using a logical relation.  

Wasm provides custom sections and the ability to implement additional type-checking by examining annotations in custom sections. We can leverage this to keep \rwasm type information around in Wasm, enabling a rich form of proof-carrying code{necula96,necula97:popl}.

Another broad area of future work is designing safe FFIs between practical typed languages that compile to RichWasm.
%There are a vast array (Cite PLDI'22 paper and Daniel's thesis).
Beyond that, we would like to support safe interop between type-safe languages compiled to RichWasm and unsafe languages such as C and C++ compiled to Wasm. The simplest solution for safe interop is to use Component Model interface types at the boundary between RichWasm (compiled to Wasm) and Wasm modules, ensuring that safe and unsafe code never mix. If we do want to support memory sharing between RichWasm and Wasm, the interop between more precisely and less precisely typed modules is reminicent of gradual typing~\cite{siek-taha06,tobin-hochstadt06,takikawa16}, but more general since the term languages have differences. This essentially requires tackling gradual typing between a language with linear capability types and one without. We plan to design a \rwasm-Wasm multilanguage, which would have to identify the static guarantees or dynamic checks we need at boundaries; then we will investigate a combination of type inference and static contract verification to eliminate these static and dynamic checks. Interop of \rwasm with MSWasm~\cite{mswasm} would be an easier way of achieving the goal since MSWasm already supports dynamic capability checking, \mbox{but to be performant, MSWasm needs specialized hardware such as CHERI~\cite{watson15}.} 
% \amal{We can remove following paragraph if we need space}
% In a \rwasm-Wasm multilanguage the \rwasm side of the boundary has type information that specifies linking requirements. If we trust these annotations, they can be used to guide program analyses of the Wasm binary on the other side of the boundary, enabling taint analysis, code debloating, and other optimizations. 

%%% Local Variables:
%%% mode: latex
%%% TeX-master: "main"
%%% End: